\documentclass[a4paper,11pt]{article}
\usepackage[toc,page]{appendix}
\usepackage{graphicx}
\usepackage{colortbl}
\usepackage{amsmath}
\usepackage{subfigure}
\usepackage{mathtools}

\DeclarePairedDelimiterX\braket[2]{\langle}{\rangle}{#1 \delimsize\vert #2}
\usepackage[utf8]{inputenc}
\usepackage{float}
\usepackage[normalem]{ulem}
\usepackage{epstopdf}
\usepackage{amsfonts,amsmath,amssymb}
\usepackage{hyperref}

\usepackage{epsfig,multicol,bbm}
\usepackage{color}
\usepackage[dvipsnames]{xcolor}

\definecolor{darkblue}{rgb}{0.0, 0.0, 0.55}
\definecolor{grey}{rgb}{0.57, 0.64, 0.69}
\definecolor{lightbrown}{rgb}{0.71, 0.4, 0.11}
\newcommand{\tcb}{\textcolor{blue}}

\newcommand{\tcr}{\textcolor{red}}

\newcommand{\be}{\begin{equation}}
\newcommand{\ee}{\end{equation}}

\date{}

\newcommand\fverb{\setbox\pippobox=\hbox\bgroup\verb}
\newcommand\fverbit{\egroup\item[\fbox{\unhbox\pippobox}]}

\newbox\pippobox
\begin{document}
\title{\bf Strongly-Coupled Anisotropic Gauge Theories and Holography in 5D Einstien-Gauss-Bonnet Gravity}
\author{S. N. Sajadi\thanks{Electronic address: naseh.sajadi@gmail.com}, 
H. R. Safari\thanks{Electronic address: hrsafari@ipm.ir}
\\
\small School of Physics, Institute for Research in Fundamental Sciences (IPM), \\P. O. Box 19395-5531, Tehran, Iran\\
}
\maketitle
\begin{quote}
{IPM/P-2023/43}
\end{quote}
\begin{abstract}
In this paper we study uncharged, non-conformal
and anisotropic systems with strong interactions
using the gauge-gravity duality by considering Einstein-Quadratic-Axion-Dilaton action in five dimension.
In fact we would like to gain insight into the influence of higher derivative gravity on the QCD system.
At finite temperature, we obtain an anisotropic black brane solution to a 5D Einstein-Gauss-Bonnet-Axion-Dilaton system. The system has been investigated and the effect of the parameter of theory has been considered. The blackening function supports the thermodynamical phase transition between small/large and AdS/large black brane for suitable parameters. We also study transport and diffusion properties, and observe in particular that the butterfly velocity that characterizes both diffusion and growth of chaos transverse to the anisotropic direction saturates a constant value in the IR which can exceed the bound given by the conformal value. We also determine the imaginary part of the heavy quark potential in a strongly coupled
plasma dual to Gauss-Bonnet gravity.
\end{abstract}
\maketitle
\section{Introduction}
Quantum Chromodynamics (QCD) is a non-abelian gauge theory that describes the behavior of the strong nuclear force. 
One of the key features of QCD is the idea of color confinement, which states that quarks and gluons cannot exist in isolation, but are always bound together to form particles such as protons and neutrons. Another important aspect of QCD is the concept of asymptotic freedom, which describes how the strong force becomes weaker at higher energies or shorter distances. The mechanism behind these two features is not well understood, and remains an active area of research. A possible way to study is gauge/gravity duality which is a duality between certain strongly interacting quantum field theories in one dimension and a weakly interacting gravitational theory in one higher dimension. 
In this framework, the extra dimension is interpreted as a scale parameter that controls the energy scale of the QCD theory\cite{Maldacena:1997re,Gubser:1998bc,Karch:2006pv,Sakai:2004cn,Klebanov:1999tb}. Quark-gluon plasma (QGP)-is believed to have existed in the early universe, in the cores of compact stars, and can also be created in high-energy heavy ion collisions in anisotropic way- is a state of matter in which, the quarks and gluons are no longer confined within the hadrons, and exist as a deconfined. {The anisotropic black hole solutions in the gauge/gravity correspondence and applications to the QGP  have been investigated in \cite{Giataganas:2017koz}-\cite{Sajadi:2023zke}.}
Higher-order gravitational models have recently received attention \cite{Sajadi:2022ybs}-\cite{Sajadi:2022tgi}, because of string theory predicts that at low energies Einstein's equations are subject to first-order corrections due to the interactions of the strings with the additional dimensions. The natural correction of Lovelock gravity to the Einstein-Hilbert gravity appears in five and higher dimensions and is given by a precise combination of quadratic curvature terms yields the second-order field equations known as the Gauss-Bonnet term \cite{Garraffo:2008hu}, \cite{Charmousis:2008kc}, \cite{Canfora:2021ttl}.
The different aspect of this theory from cosmology to black hole solutions have been studied in \cite{Deruelle:2003ck}-\cite{Padmanabhan:2013xyr}. {In this paper, we extend the work of \cite{Giataganas:2017koz} to the Einstein Quadratic Gravity, which is general relativity extended by quadratic curvature invariant in the action to find the effect of higher derivative terms on QGP, with the difference that in \cite{Giataganas:2017koz}, they were fixed $V (\phi)$ and $Z(\phi)$, which were chosen in order to obtain specific properties in the IR and UV. In the present paper, however, we fix both $V (\phi)$ and $Z(\phi)$ from the equations of motion, because the functionality of these potentials in terms of the couplings of QGP is unknown to us.}
\\
The paper is organized as follows. In section \ref{sec2} we construct the anisotropic $5$-dimensional solution with an arbitrary dynamical exponent, an exponential quadratic warp function, in the framework of EGB gravity. We obtained the approximately solution for blackening function and other unknown quantities up to the first order of the theoretical parameter. We have shown the behavior of the quantities with plots and we discuss the thermodynamics of the constructed background in section \ref{sec3}. In section \ref{sec4}, we obtain transport and diffusion properties in anisotropic theories and observe in particular that the butterfly velocity that characterizes both diffusion and growth of chaos transverse to the anisotropic direction saturates a constant value in the IR which can exceed the bound given by the conformal value.
In section \ref{sec5}, the imaginary part of the quark-antiquark potential in different directions. 
In section \ref{sec6}, we obtained the Jet quenching related to the suppression of high-energy jets of particles produced in high energy heavy-ion collisions due to their interaction with QGP. We finish the paper with some concluding remarks in section \ref{conclud}.

\section{Basic Formalism}\label{sec2}
We consider a $5$-dimensional Einstein-Quadratic-Axion-Dilaton action as follows
\begin{equation}\label{action1}
S=\dfrac{1}{16\pi G_{5}}\int d^{5}x \sqrt{-g}L,
\end{equation}
where the Lagrangian is
\begin{equation}
L=R+\gamma R_{a b c d}R^{a b c d}+\beta R_{a b}R^{a b}+\alpha R^{2}-\dfrac{1}{2}\partial_{\mu}\phi\partial^{\mu}\phi +V(\phi)-\dfrac{1}{2}Z(\phi)\partial_{\mu}\chi\partial^{\mu}\chi,
\end{equation}
and $V(\phi)$ is the potential energy for the dilaton field $\phi$, $Z(\phi)$ is the coupling $\phi$ to the axion field $\chi$. $G_{5}$ is the Newton constant in five dimensions and $\alpha$, $\beta$ and $\gamma$ are coupling constants of theory. The Einstein and Quadratic terms describe the gravitational interaction between matter in QCD, here the matter is quarks, gluons and plasma. The dilaton field modifies the running of the QCD coupling constant which reflected the non conformality of QCD in gravity. The axion, which is dual to the gauge theory $\theta$-term (is a term in the QCD Lagrangian that describes the possibility of CP violation in strong interactions), is responsible for inducing the anisotropy \cite{Gursoy:2007er},\cite{Gursoy:2007cb}. 
The variation of the action (\ref{action1}) over metric $g_{\mu \nu}$, the scalar field $\phi$ and $\chi$ gives the field equations as follows
\begin{small}
\begin{eqnarray}
&&E_{\mu\nu}=G_{\mu \nu}+\alpha\left[2R(R_{\mu\nu}-\frac{1}{4}g_{\mu\nu}R)+2(g_{\mu\nu}\square-\nabla_{\mu}\nabla_{\nu})R\right]+\beta\left[(g_{\mu\nu}\square-\nabla_{\mu}\nabla_{\nu})R+\square G_{\mu\nu}+2R^{\lambda \rho}(R_{\mu\lambda\nu\rho}\right.  \nonumber\\
&&\left.-\frac{1}{4}g_{\mu\nu}R_{\lambda\rho})\right]
+\gamma\left[-\frac{1}{2}g_{\mu\nu}R_{\alpha\beta\gamma\eta}R^{\alpha\beta\gamma\eta}+
2R_{\mu\lambda\rho\sigma}R_{\nu}{}^{\lambda\rho\sigma}+4R_{\mu\lambda\nu\rho}R^{\lambda\rho}-4R_{\mu\sigma}
R^{\sigma}_{\nu}+4\square R_{\mu\nu}-2\nabla_{\mu}\nabla_{\nu}R\right]\nonumber\\
&&=\dfrac{1}{2}\nabla_{\alpha}\phi\nabla_{\beta}\phi+\dfrac{1}{2}g_{\alpha\beta}V(\phi)-\dfrac{1}{4}g_{\alpha\beta}\nabla_{\gamma}\phi\nabla^{\gamma}\phi+\dfrac{1}{2}Z(\phi)\nabla_{\alpha}\chi\nabla_{\beta}\chi -\dfrac{1}{4}g_{\alpha\beta}Z(\phi)\nabla_{\gamma}\chi\nabla^{\gamma}\chi ,\label{eqq3} \\
&&\square\phi -\dfrac{1}{2}\nabla_{\alpha}\chi \nabla^{\alpha}\chi Z^{\prime}(\phi)+V^{\prime}(\phi)=0,\\
&&Z(\phi)\square \chi+\nabla_{\alpha}\phi\nabla^{\alpha}\chi Z^{\prime}(\phi)=0,
\end{eqnarray}
\end{small}
\noindent where $G_{\mu\nu}$ is the Einstein tensor. We use the metric ansatz $g_{\mu \nu}$ in the following form:
\begin{equation}
ds^{2}=A^{2}(z)\left(-g(z)dt^{2}+\dfrac{dz^2}{g(z)}+h^{2}(z)dx^{2}+dy_{1}^{2}+dy_{2}^{2}\right),\;\;\;\phi=\phi(z)
\end{equation}
where $A$, $g$ and $h$ are functions of the holographic coordinate $z$. The warp function $A(z)$ is related to the energy scale of QCD by compactification of extra dimension. $g(z)$ is the blackening function which is dual to thermal states in the QCD. Also, $z=0$ and $z \to \infty$ correspond to boundary and anisotropic IR region, respectively. 
 Isotropy in the $y_{1}y_{2}$-directions is respected, but not in the $x$-direction unless $h(z) = 1$.
Using the ansatz of the metric, it is easy to obtain the equations of motion for the background fields.
The field equations for $\chi$ and $\phi$ are given by:
\begin{align}
&\;\;\;\;\;\;\;\;\;\;\;\;\;\;\;\;\;\;\;\;\;\;\;\;\;\;\;\;\;\;\;\;\;\;\;\;\;\;\;\;\;\;\;\;\;\dfrac{d^{2}\chi(x)}{dx^{2}}=0,\label{eqqchi}\\
&2h^2
A^{3}\dfrac{dV}{d\phi}-A\dfrac{dZ}{d\phi}
 \left({\frac {d\chi}{dx}}\right) ^{2}+6h^2g\phi^{\prime}A^{\prime}+2Agh\phi^{\prime}h^{\prime} +2Ah^2g\phi^{\prime\prime} +2Ah^2\phi^{\prime} g^{\prime}=0,\label{eqqphi}
\end{align}
where prime is differential with respect to $z$. From solving equation \eqref{eqqchi}, we have:
\begin{equation}\label{eqqchs}
\chi(x)=c_{1}x+c_{2}.
\end{equation}
For $\beta=-4\alpha$, $\gamma=\alpha$ (Einstien-Gauss-Bonnet Gravity) and using \eqref{eqqchs} we get the other components of the field equation which was presented in appendix \eqref{app2}.

{We obtain the $Z(\phi)$ from $E_{y_{1}y_{2}}$ and $\phi^{\prime}$ from $E_{zz}$. Inserting $Z(\phi)$ into the $E_{xx}$ we obtained $V(\phi)$. Finally, inserting $Z(\phi)$, $V(\phi)$ and $\phi^{\prime}$ into the $E_{tt}$ a constrained differential equation obtains as follows}
\begin{eqnarray}\label{eqfe}
&&2hA^2g^{\prime\prime}+2A^{2}g^{\prime}h^{\prime}+6Ahg^{\prime}A^{\prime}+\dfrac{8\alpha}{A^{3}}\left[-hgAg^{\prime\prime}A^{\prime 2}-gA^{2}h^{\prime}g^{\prime}A^{\prime\prime}-A^{2}gA^{\prime}h^{\prime}
g^{\prime\prime}+8ghg^{\prime}A^{\prime 3}\right.\nonumber\\
&&\left. -2ghAA^{\prime}g^{\prime}A^{\prime} -hAA^{\prime 2}g^{\prime 2}-gA^{2}A^{\prime}g^{\prime}h^{\prime\prime}-A^{2}h^{\prime}A^{\prime}g^{\prime 2}-Agh^{\prime}g^{\prime}A^{\prime 2}\right]=0.
\end{eqnarray}
To solve the above equation in IR region, we assume
\begin{equation}\label{eqmetric}
ds^{2}=(cz)^{k_{1}-2}\left[-g(z)dt^{2}+\dfrac{dz^{2}}{g(z)}+dy_{1}^{2}+dy_{2}^{2}+\dfrac{\mathfrak{c}dx^{2}}{(cz)^{k_{0}-2}}\right].
\end{equation}
For $k_{1}=0$ the solution exhibits a Lifshitz-like scaling, and for $k_{1}\neq 0$ the metric \eqref{eqmetric}  has the hyperscaling violation property. The case $k_{0}=2$ and $\mathfrak{c}=1$ corresponds to isotropic solution.
{Using \eqref{eqmetric}, the differential equation \eqref{eqfe} becomes}
\begin{small}
\begin{align}\label{eqqdif}
z^{2}(cz)^{k_{1}-2}\left[zg^{\prime\prime}-\dfrac{1}{2}(k_{0}-3k_{1}+4)g^{\prime}\right]+\alpha (k_{1}-2)(k_{0}-k_{1})\left[zgg^{\prime\prime}+g^{\prime}(zg^{\prime}-\dfrac{1}{2}g(k_{0}-k_{1}+4))\right]=0.
\end{align}
\end{small}
To solve \eqref{eqqdif} with the generic constants $k_{0}$ and $k_{1}$, we work in a perturbative manner as follows
\begin{equation}\label{eqqg}
g(z)=g_{0}(z)+\alpha g_{1}(z)+\mathcal{O}(\alpha^2),
\end{equation}
therefore, one can obtain $g_{0}$ and $g_{1}$ as follows
\begin{align}
g_{0}(z)=1-\left(\dfrac{z}{z_{h}}\right)^{3+\frac{k_{0}-3k_{1}}{2}}
\end{align}
and
\begin{small}
\begin{align}
&g_{1}(z)=1-\left(\dfrac{z}{z_{h}}\right)^{3+\frac{k_{0}-3k_{1}}{2}}+\dfrac{(k_{1}-2)(k_{0}-k_{1})(k_{0}-3k_{1}+6)}{k_{0}+6-5k_{1}}c^{2}(cz)^{-k}\left(\dfrac{z}{z_{h}}\right)^{3+\frac{k_{0}-3k_{1}}{2}}-\nonumber\\
&\dfrac{(k_{1}-2)(k_{0}-k_{1})(k_{0}+6-3k_{1})^{2}c^{-k_{1}+2}}{2(k_{0}+6-5k_{1})(k_{0}+6-4k_{1})}\dfrac{z^{3+\frac{k_{0}-3k_{1}}{2}}}{z_{h}^{3+\frac{k_{0}-k_{1}}{2}}}-\dfrac{(k_{1}-2)(k_{0}-k_{1})(k_{0}+6-3k_{1})c^{-k_{1}+2}}{2(k_{0}+6-4k_{1})}\dfrac{z^{6+k_{0}-4k_{1}}}{z_{h}^{k_{0}-3k_{1}+6}}
\end{align}
\end{small}
here, we imposed the following conditions on the metric
\begin{equation}\label{eqqcond}
 g(0)=1,\;\;\;\; g(z_{h})=0.
\end{equation}
\begin{figure}[H]\hspace{0.4cm}
\centering
\subfigure[$\alpha=0.2,z_{h}=1,k_{1}=-2,c=1$]{\includegraphics[width=0.45\columnwidth]{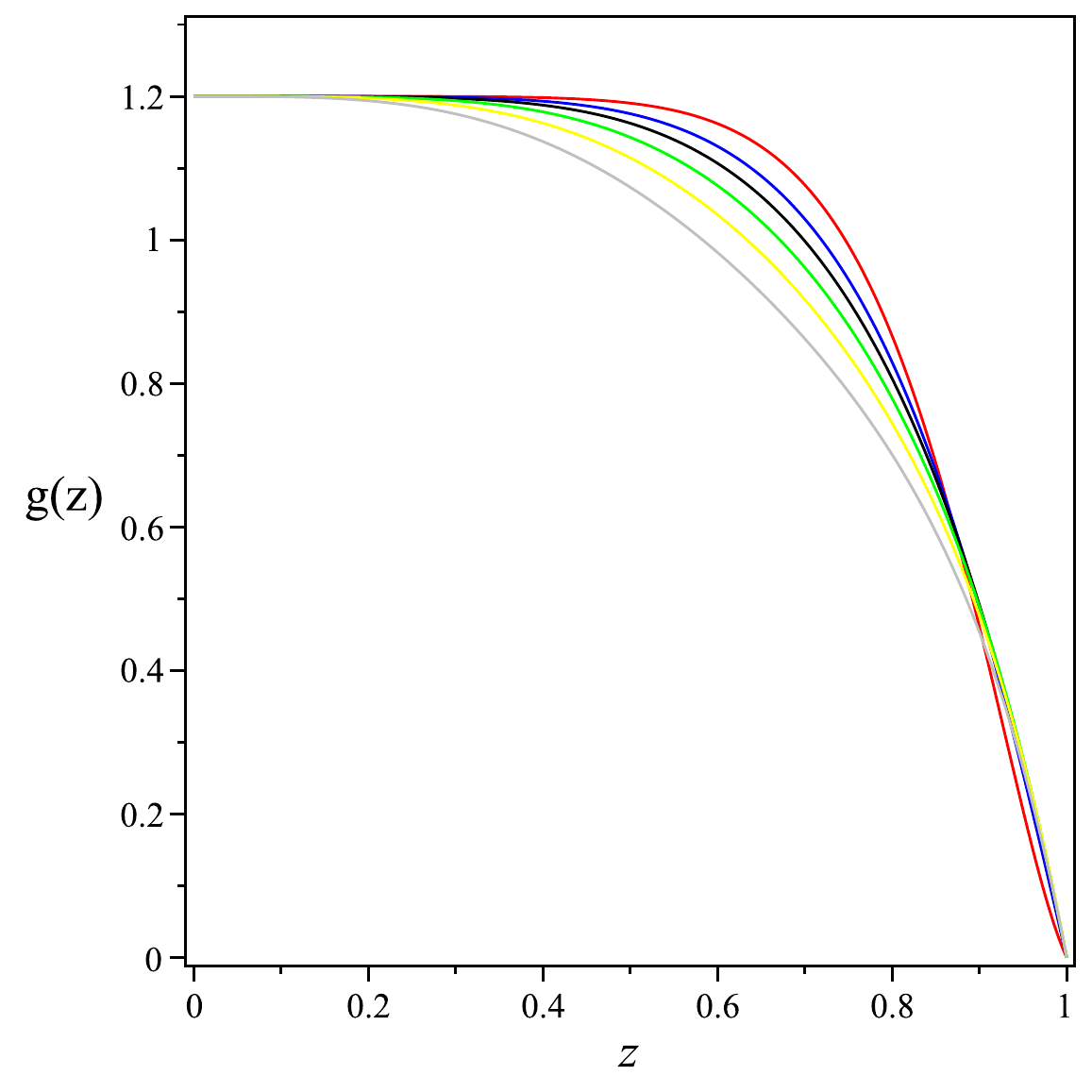}}
\subfigure[$\alpha=0.2,z_{h}=1,k_{1}=-2,k_{0}=3,c=1$]{\includegraphics[width=0.45\columnwidth]{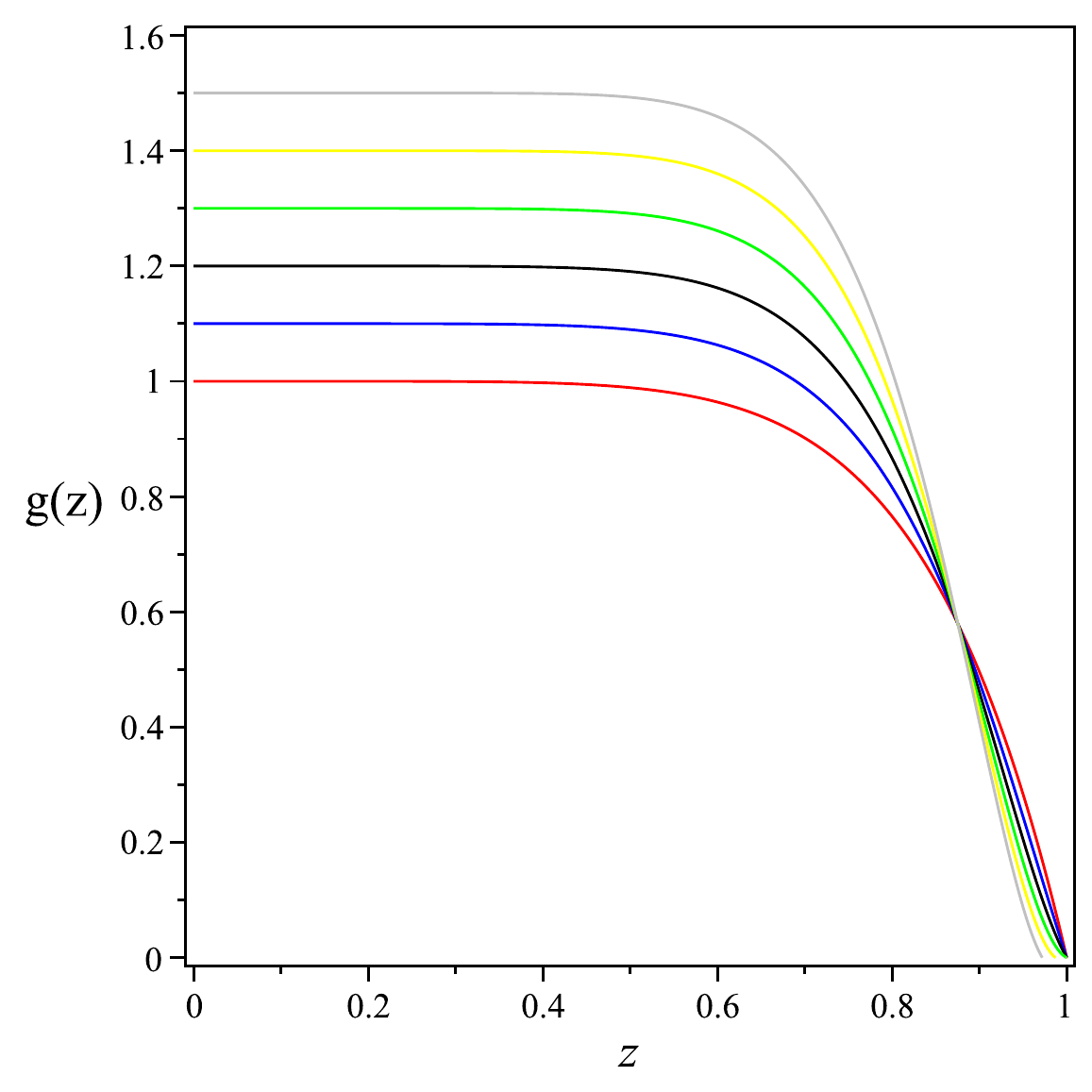}}
\caption{Plots of $g(z)$ in terms of $z$ for $k_{0}=\textcolor{red}{1},\textcolor{blue}{-1},\textcolor{black}{-2},\textcolor{green}{-3},\textcolor{yellow}{-4},\textcolor{gray}{-5}$ (left) and for $\alpha=\textcolor{red}{0},\textcolor{blue}{0.1},\textcolor{black}{0.2},\textcolor{green}{0.3},\textcolor{yellow}{0.4},\textcolor{gray}{0.5}$ (right).} 
\label{gzplote1}
\end{figure}
{The solutions we found do not interpolate between AdS in UV and anisotropic in IR. These solutions have fixed $c$-function \cite{Chu:2019uoh} in contrast to the full solutions used in \cite{Giataganas:2017koz}.}
The behavior of the metric function is depicted in Fig.(\ref{gzplote1}). The main feature is that the metric function values decrease faster for larger $\alpha$ and $k_{0}$. In the isotropic case {($k_{0}=2$)} the metric function values are larger than in the anisotropic ones {($k_{0}\neq 2$)} (Fig. \ref{gzplote1}a).  By inserting (\ref{eqqg}) in $E_{zz}$, one can obtain $\phi$ for $\alpha\ll 1$ as follows:
\begin{align}
\phi(z)=&\dfrac{1}{2}\sqrt{6k_{1}^{2}-12k_{1}+4k_{0}-2k_{0}^{2}}\ln\left(\dfrac{z}{z_{h}}\right)+\dfrac{\alpha(k_{1}-2)^{2}(k_{0}+3k_{1}-2)(k_{1}-k_{0})c^{2-k_{1}}z_{h}^{-k_{1}}}{2\sqrt{2}k_{1}(k_{0}+6-5k_{1})\sqrt{3k_{1}^{2}-6k_{1}-k_{0}^{2}+2k_{0}}}\nonumber\\
&\left[3k_{1}-6-k_{0}+(k_{0}+6-5k_{1})\left(\dfrac{z}{z_{h}}\right)^{-k_{1}}+2k_{1}\left(\dfrac{z}{z_{h}}\right)^{3+\frac{k_{0}-5k_{1}}{2}}\right]+\mathcal{O}(\alpha^2),
\end{align}
here, we imposed the condition $\phi(z_{h}) = 0$.

\begin{figure}[H]\hspace{0.4cm}
\centering
\subfigure[$z_{h}=1,\alpha=0.2,k_{1}=-2,c=1$]{\includegraphics[width=0.4\columnwidth]{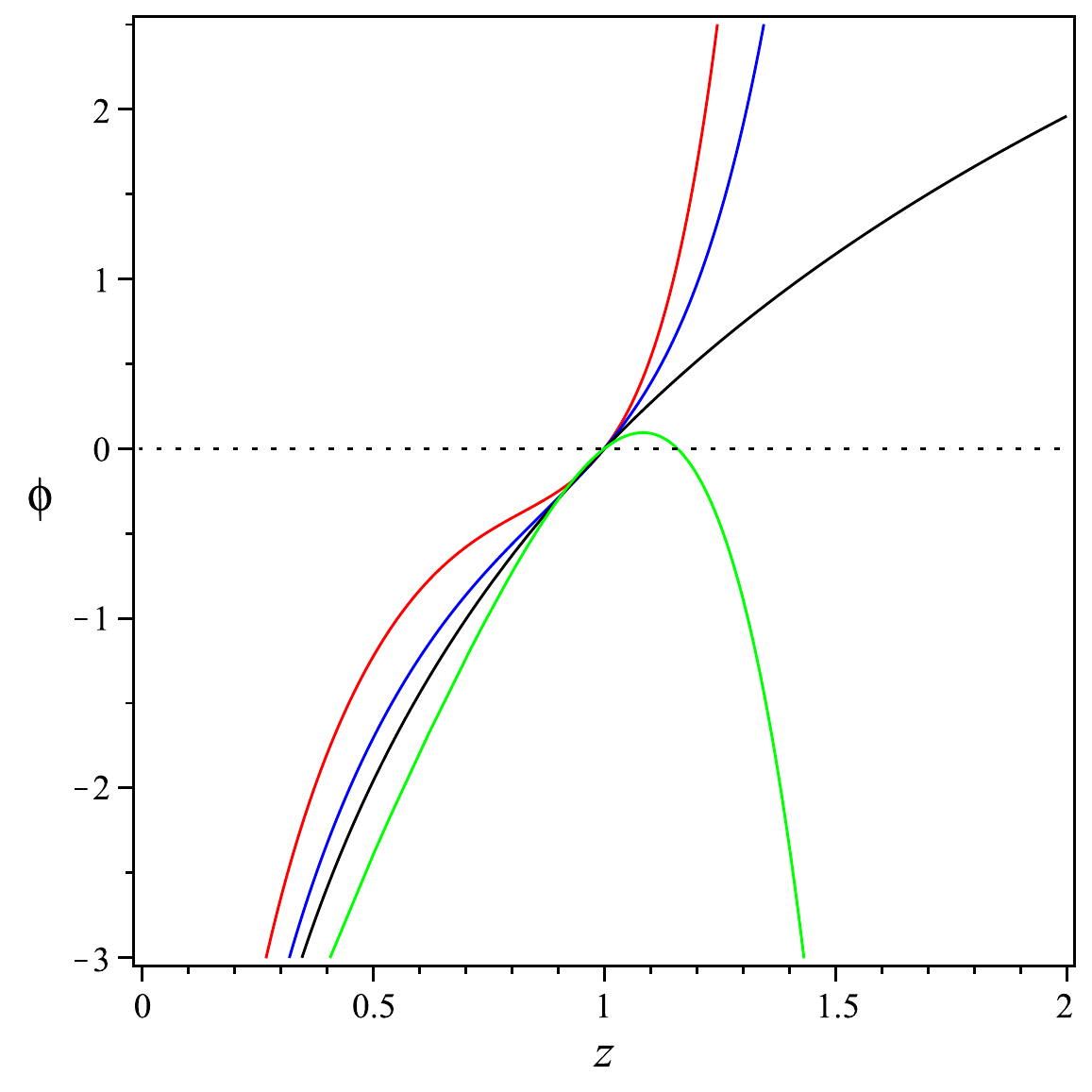}}
\subfigure[$z_{h}=1,k_{0}=1,k_{1}=-2,c=1$]{\includegraphics[width=0.4\columnwidth]{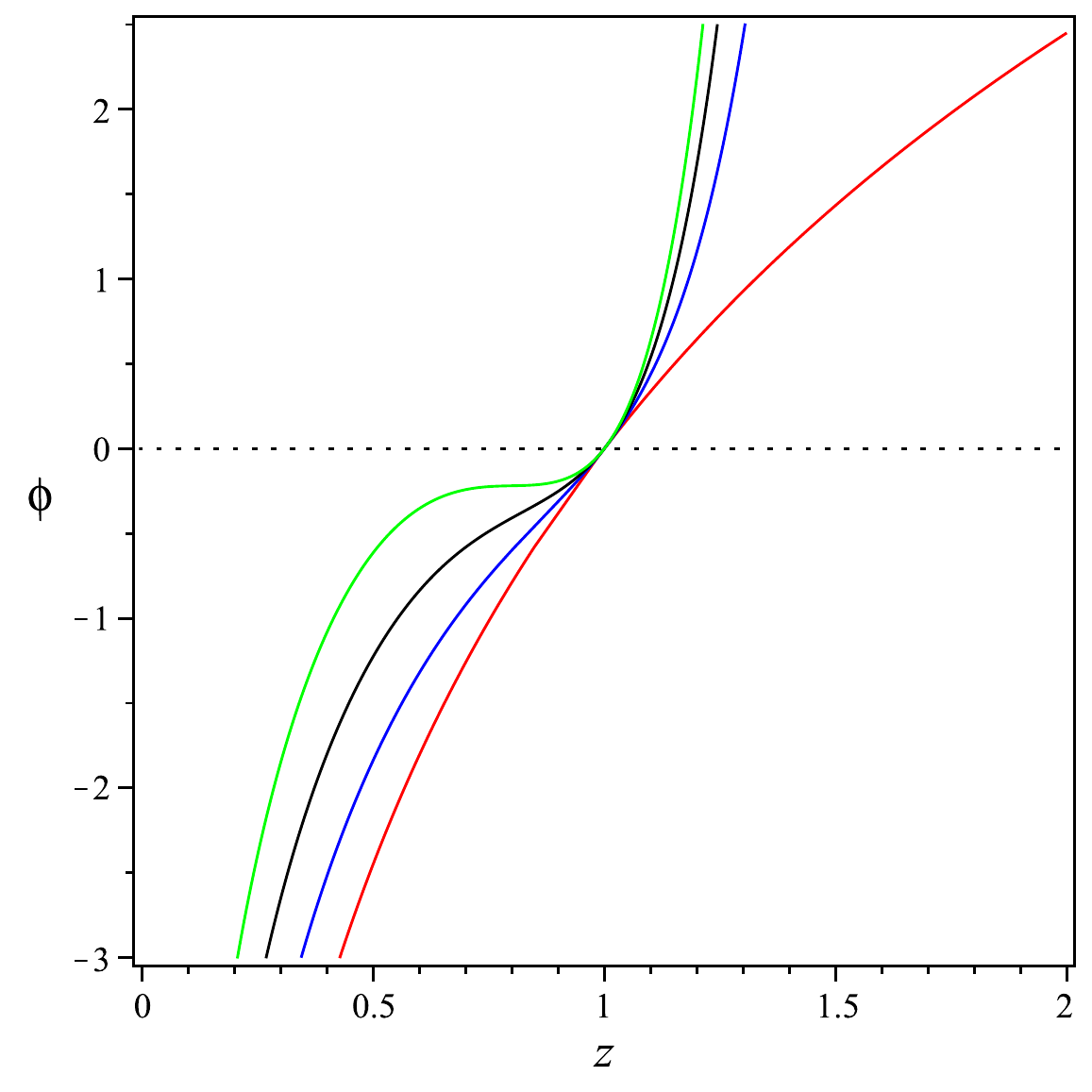}}
\caption{Plot of $\phi$ in terms of $z$ for $k_{0}=\textcolor{red}{1},\textcolor{blue}{-1},\textcolor{black}{-2},\textcolor{green}{-3}$ (left) and for $\alpha=\textcolor{red}{0},\textcolor{blue}{0.1},\textcolor{black}{0.2},\textcolor{green}{0.3}$ (right).} 
\label{phiplote}
\end{figure}
In figure (\ref{phiplote}), the scalar field in terms of $z$ for
different values of parameters have been shown. As can be seen the imaginary part of scalar
field inside and outside the black brane has a zero value and is stable. By increasing $k_{0}$ and $\alpha$ in $0 < z < z_{h}$, the scalar field increase. 
By inserting (\ref{eqqg}) in $E_{y_{1}y_{1}}$, one can obtain $Z(\phi)$ for $\alpha\ll 1$ as follows:
\begin{align}\label{zphi}
{Z(\phi)=}&\dfrac{1}{2c_{1}}(k_{0}-2)(3k_{1}-k_{0}-6)c^{2-k_{0}}z_{h}^{-k_{0}}e^{\frac{-2\phi k_{0}}{\sqrt{3k_{1}^{2}-6k_{1}-k_{0}^{2}+2k_{0}}}}+\alpha \hat{Z}+\mathcal{O}(\alpha^2),
\end{align}
{$\hat{Z}$ provided in appendix (\ref{app3})}.

\begin{figure}[H]\hspace{0.4cm}
\centering
\subfigure{\includegraphics[width=0.45\columnwidth]{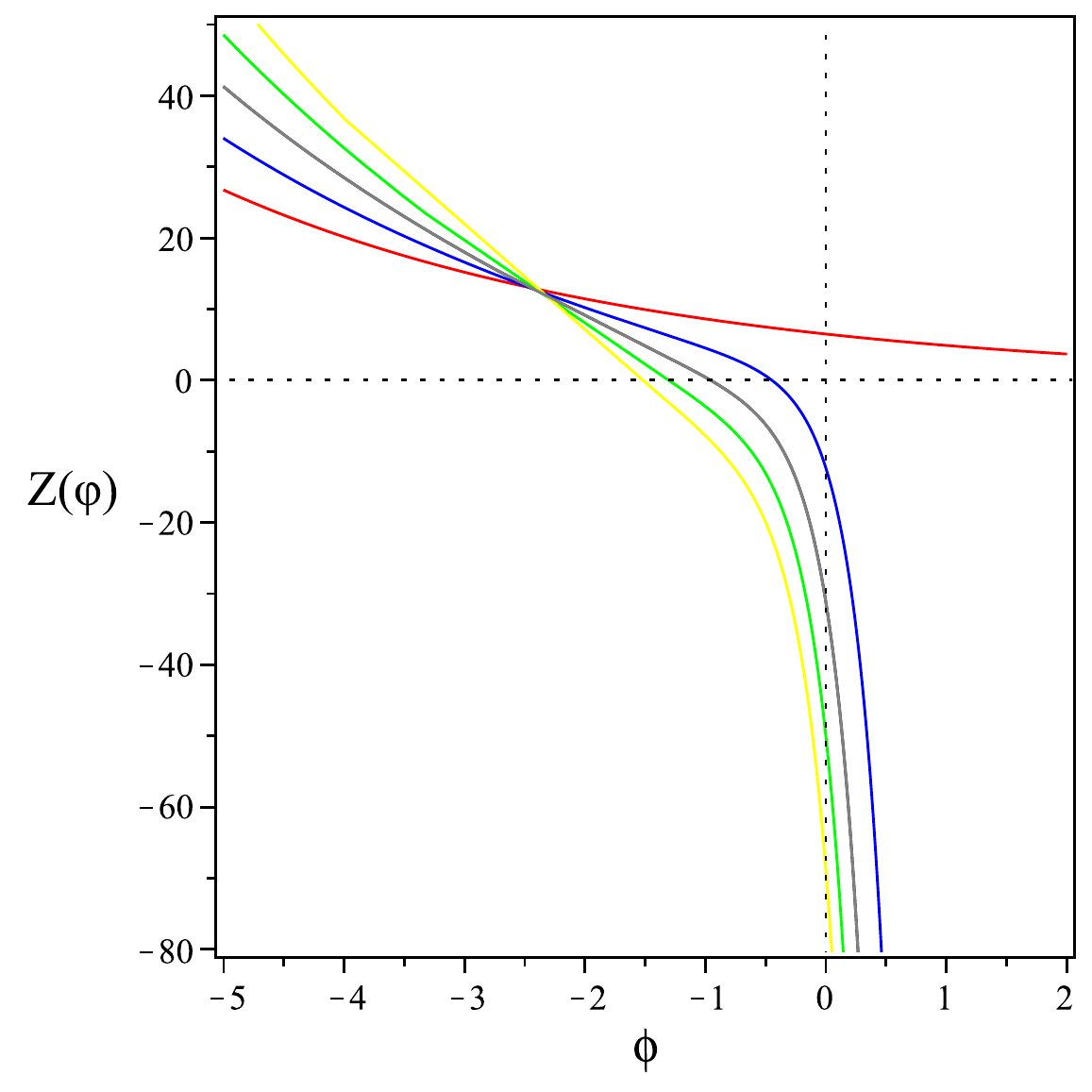}}
\subfigure{\includegraphics[width=0.45\columnwidth]{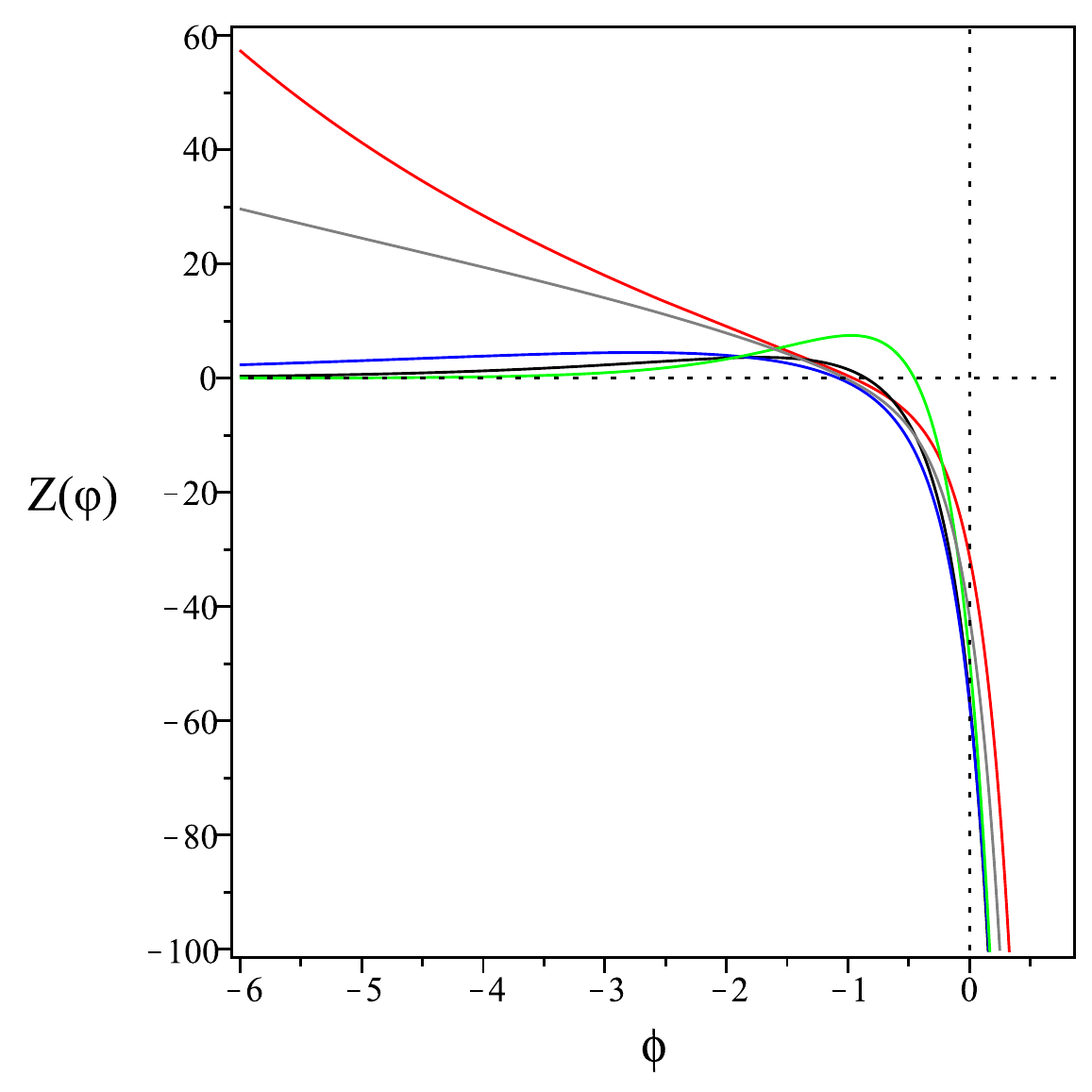}}
\caption{Plot of $Z(\phi)$ in terms of $\phi$ for $k_{0}=\textcolor{red}{1},\textcolor{blue}{-1},\textcolor{black}{-2},\textcolor{green}{-3},k_{1}=-2,\alpha=0.2,c=1$ (left) and for $k_{0}=1,\alpha=\textcolor{red}{0},\textcolor{blue}{0.1},\textcolor{black}{0.2},\textcolor{green}{0.3},\textcolor{yellow}{0.4}$(right).} 
\label{Zpplote}
\end{figure}
\begin{figure}[H]\hspace{0.4cm}
\centering
\subfigure[$z_{h}=1,\alpha=0.2,k_{1}=-2,c=1$]{\includegraphics[width=0.45\columnwidth]{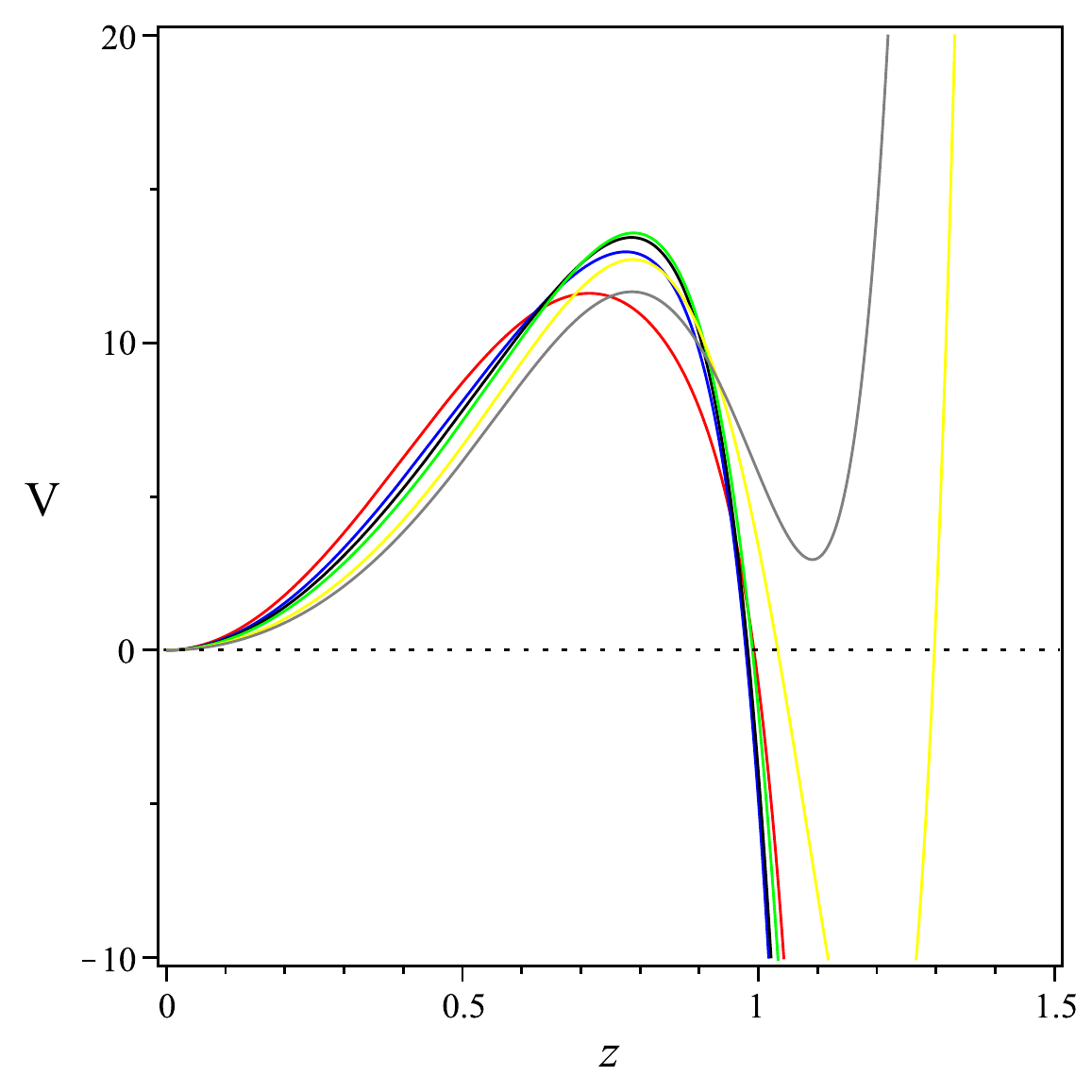}}
\subfigure[$z_{h}=1,k_{0}=1,k_{1}=-2,c=1$]{\includegraphics[width=0.45\columnwidth]{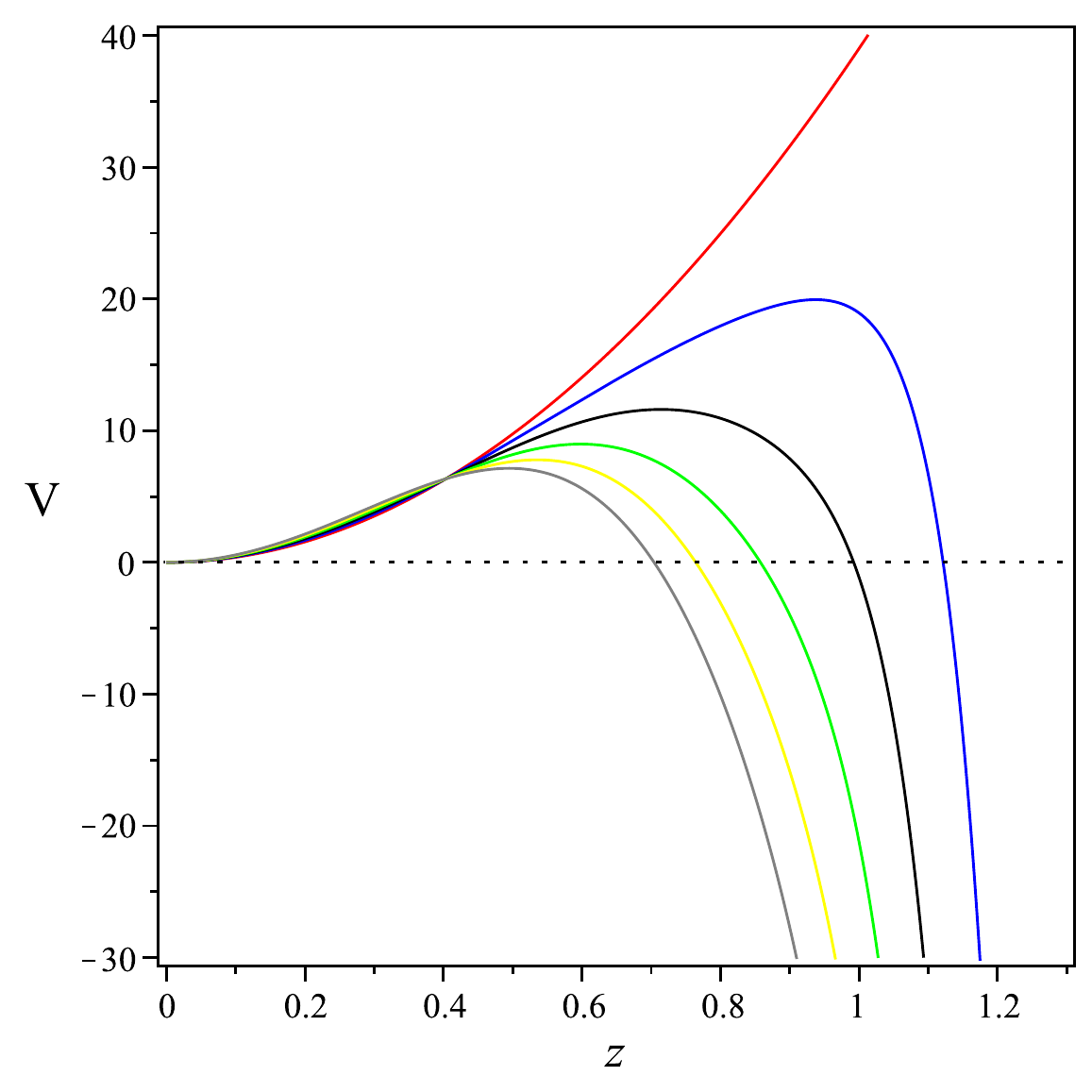}}
\caption{Plot of $V$ in terms of $z$ for $k_{0}=\textcolor{red}{1},\textcolor{blue}{-1},\textcolor{black}{-2},\textcolor{green}{-3}$,$\textcolor{yellow}{-4}$,$\textcolor{gray}{-5}$ (left) and for $\alpha=\textcolor{red}{0},\textcolor{blue}{0.1},\textcolor{black}{0.2},\textcolor{green}{0.3},\textcolor{yellow}{0.4},\textcolor{gray}{0.5}$ (right).} 
\label{Vpplote}
\end{figure}

In figure (\ref{Zpplote}), the behavior of $Z(\phi)$ in terms of $\phi$ are shown. As $\alpha$ increases and $k_{0}$ decreases, $Z$ decrease.
Finally from $E_{xx}$, one can get $V(\phi)$. In the case of $\alpha\ll 1$, one can get
\begin{align}\label{Vphi}
V(\phi)=-\dfrac{3}{4}(k_{1}-2)(k_{0}+6-3k_{1})c^{2-k_{1}}z_{h}^{-k_{1}}e^{\frac{-2k_{1}\phi}{\sqrt{6k_{1}^{2}+4k_{0}-2k_{0}^{2}-12k_{1}}}}+\alpha \hat{V}+\mathcal{O}(\alpha^2),
\end{align}
{$\hat{V}$ provided in appendix \eqref{app3}.}
{As can be seen the leading term of potential ($\alpha=0$) is exponential which shows that we are in IR regime and the solution is not asymptotically AdS.}
In Fig. (\ref{Vpplote}), the behavior of scalar potential in terms of $z$ for finite $\alpha$ has been shown. {As can be seen from the plots for $\alpha\neq 0$, it is clear there are multiple minima, and the solution can flow in the UV, yielding AdS asymptotics}. In
the both panels, between $0 < z < z_{h}$ by increasing $k_{0}$ and $\alpha$, the potential increase. For the case of $k_{1}=0$ and $k_{0}=2+\epsilon$, the potential becomes
 \begin{equation}
 V\sim 12c^{2}+12c^{2}(1-2c^{2})\alpha+\dfrac{3}{2}\epsilon c^{2}+\mathcal{O}(\epsilon\alpha).
 \end{equation}

\subsection{Thermodynamics of the background}\label{sec3}
\begin{figure}[H]\hspace{0.4cm}
\centering
\subfigure[$\alpha=0.2,k_{1}=-2,c=1$]{\includegraphics[width=0.3\columnwidth]{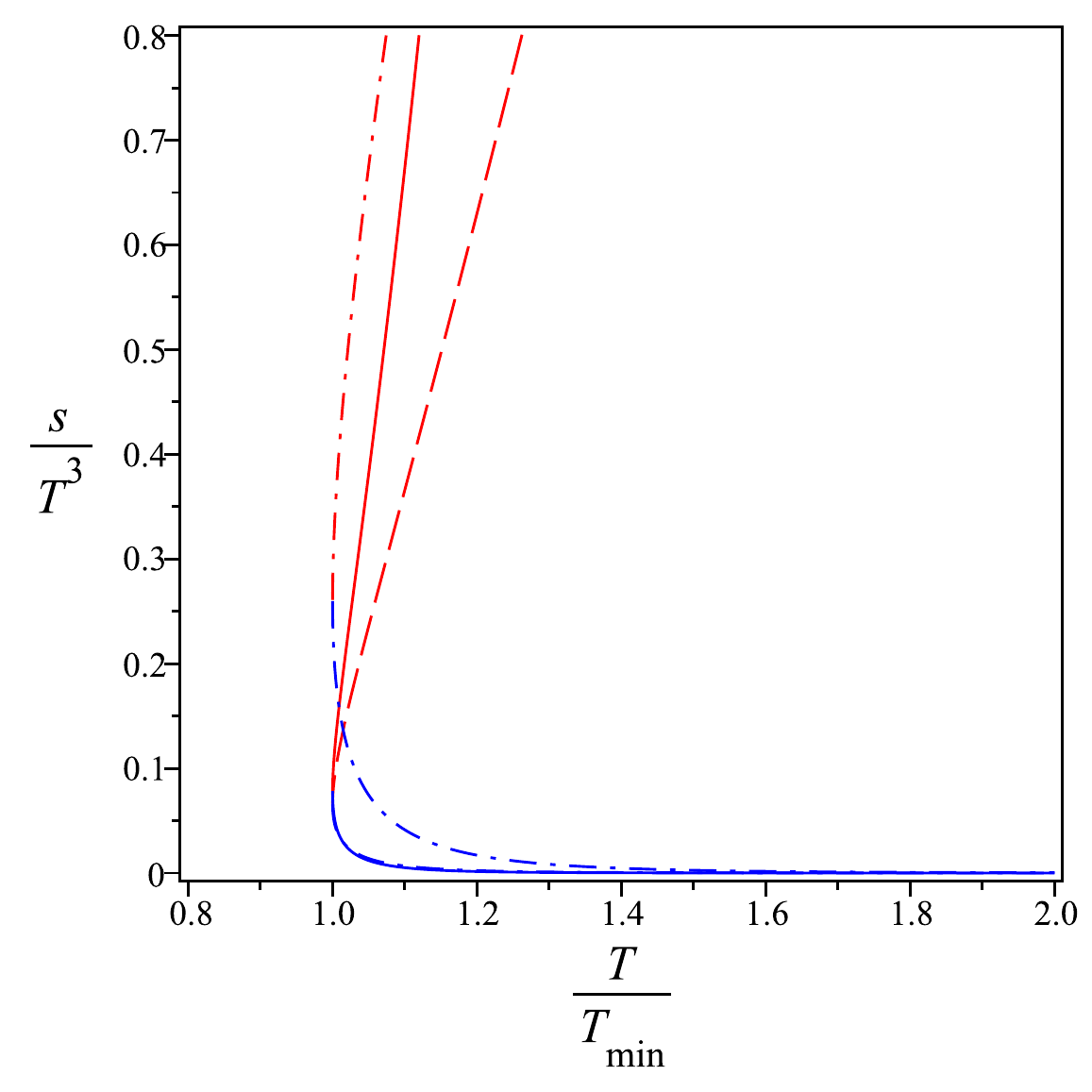}}
\subfigure[$\alpha=0.2,k_{1}=-2,c=1$]{\includegraphics[width=0.3\columnwidth]{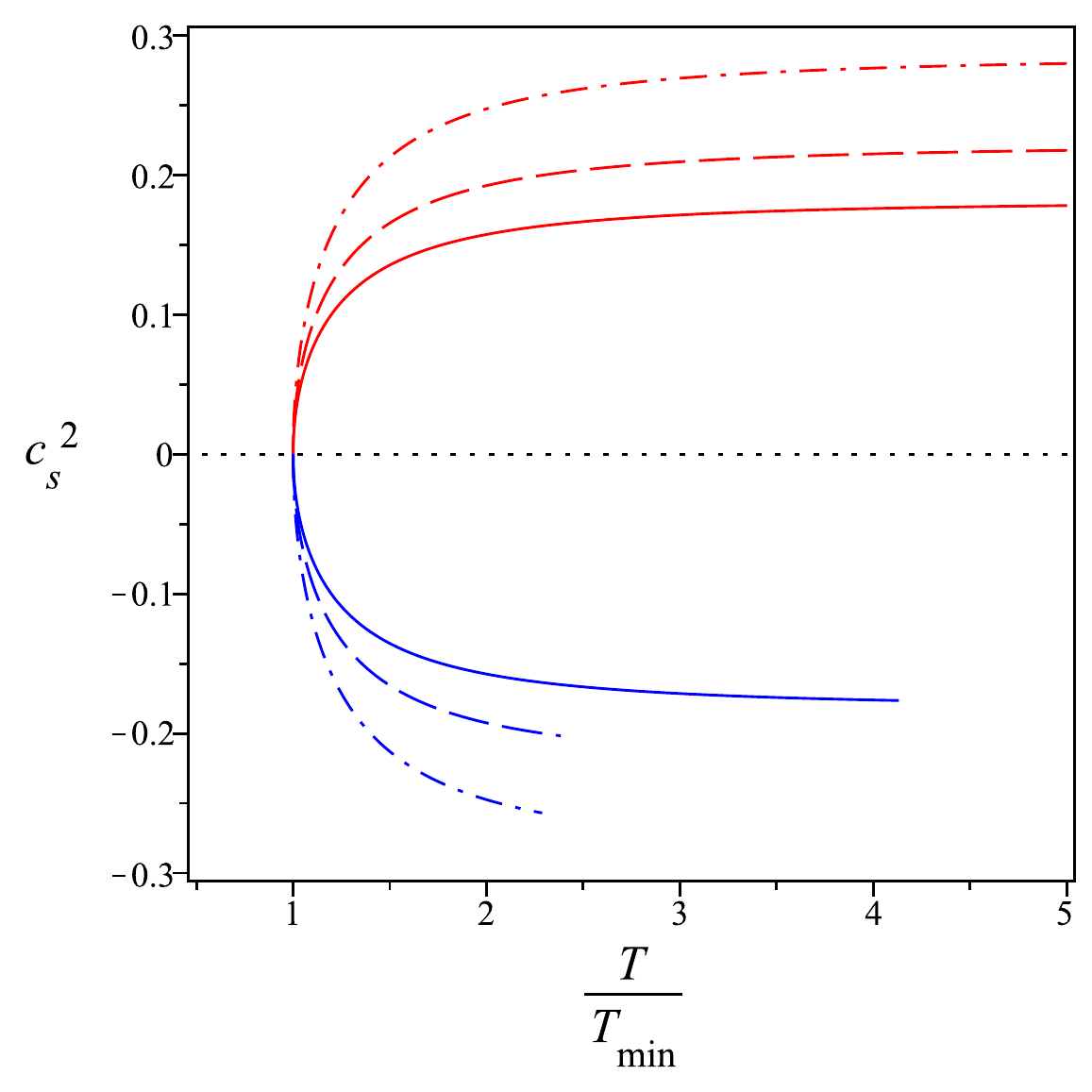}}
\subfigure[$\alpha=0.2,k_{1}=-2,c=1$]{\includegraphics[width=0.3\columnwidth]{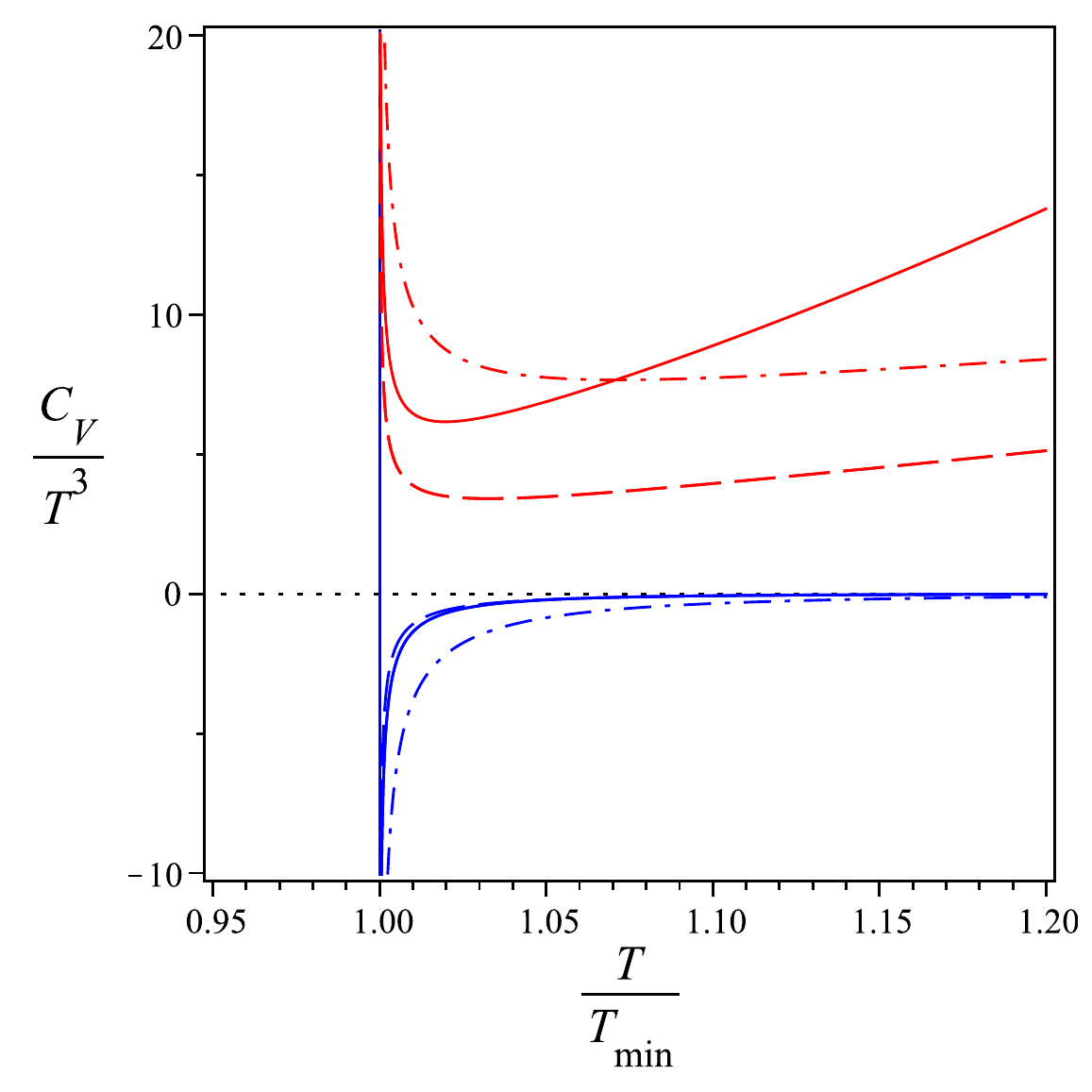}}
\caption{Plot of $s/T^3$ (left), $c_{s}^{2}/T^{3}$(middle) and $C_{V}/T^{3}$ (right) in terms of $T/T_{min}$ for $k_{0}=1,-1,-3$.} 
\label{szhplote}
\end{figure}

\begin{figure}[H]\hspace{0.4cm}
\centering
\subfigure{\includegraphics[width=0.45\columnwidth]{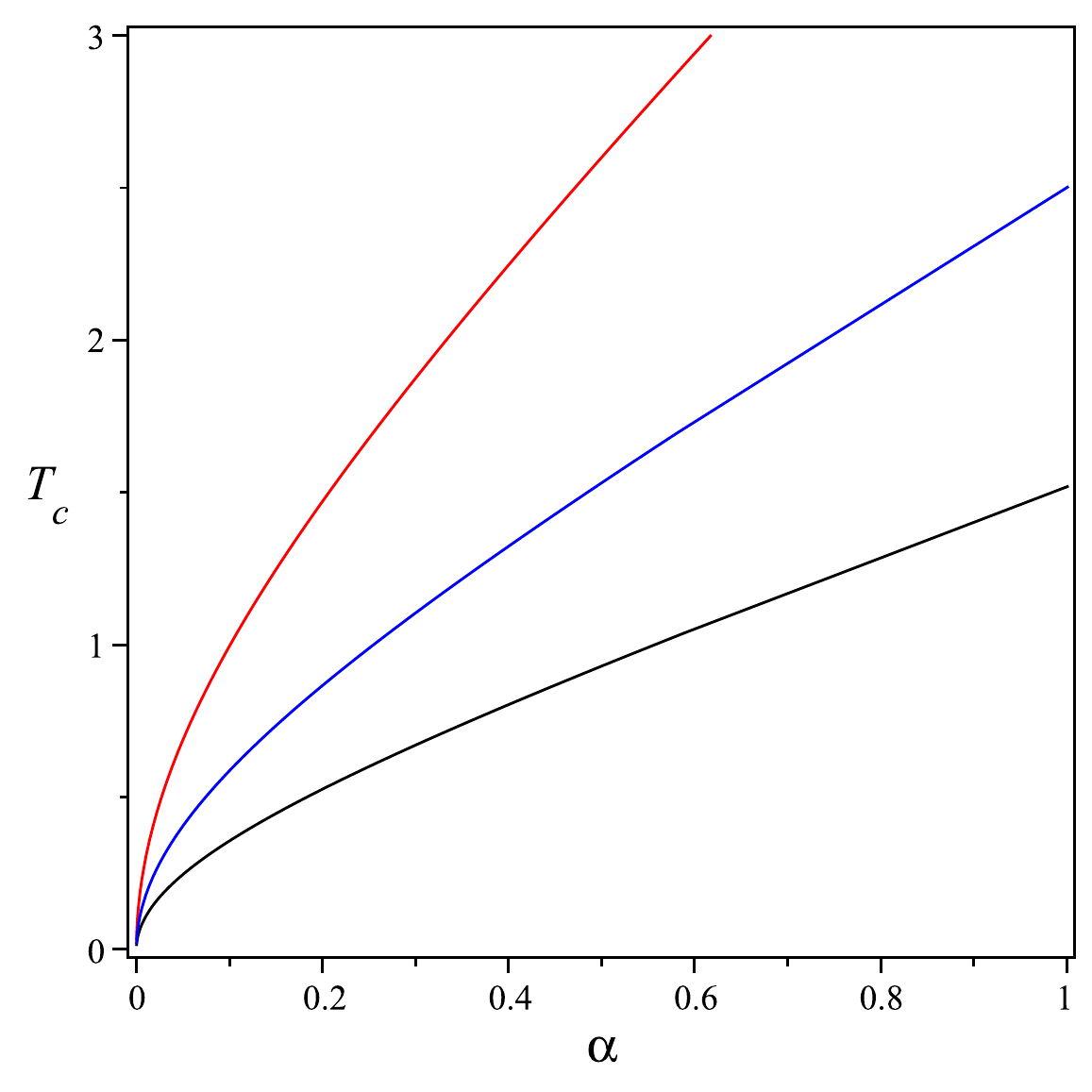}}
\caption{Plot of $T_{c}$ in terms of $\alpha$ for $c=1$, $(\textcolor{red}{k_{0}={2},k_{1}=0})$, $(\textcolor{blue}{k_{0}={0},k_{1}=0})$ and $(\textcolor{black}{k_{0}={-4},k_{1}=-2})$.} 
\label{Tzhplote1}
\end{figure}

In this subsection, we explore the thermodynamics of the black hole solution (\ref{eqqg}). In order to investigate the thermodynamic properties of the black hole we need to obtain some relevant thermodynamic quantities. The temperature of the black hole is obtained as follows:
\begin{equation}\label{temp}
T=\left\vert\dfrac{g^{\prime}}{4\pi}\right\vert =\dfrac{\vert k_{0}+6-3k_{1}\vert}{8\pi z_{h}}+\alpha\left[\dfrac{1}{8\pi z_{h}}+\dfrac{c^{-k_{1}+2}(k_{0}+6-3k_{1})^{3}(k_{1}-2)(k_{0}-k_{1})}{16\pi z_{h}^{k_{1}+1}(k_{0}+6-5k_{1})(k_{0}+6-4k_{1})}\right]+\mathcal{O}(\alpha^{2}).
\end{equation}
The temperature monotonically decreases with the increase of the horizon if $k_{1}<-2$. For $k_{1}\geq -2$, the temperature has a minimum. In Fig. (\ref{Tzhplote})a, the variation of temperature with respect to the horizon radius $z_{h}$ for different values of $k_{0}$ is shown. As can be seen there exists a minimum temperature $T_{min}$ below which no black hole solution exist (thermal gas). However, for $T > T_{min}$, there are two black hole solutions, a large and a small one (deconfined quark gluon plasma phase). The small black hole phase for which $T$ increases with $z_{h}$ whereas the large black hole phase for which $T$ decreases with $z_h$. For $\alpha=0$, one can get the result of \cite{Giataganas:2017koz}.
\begin{figure}[H]\hspace{0.4cm}
\centering
\subfigure{\includegraphics[width=0.3\columnwidth]{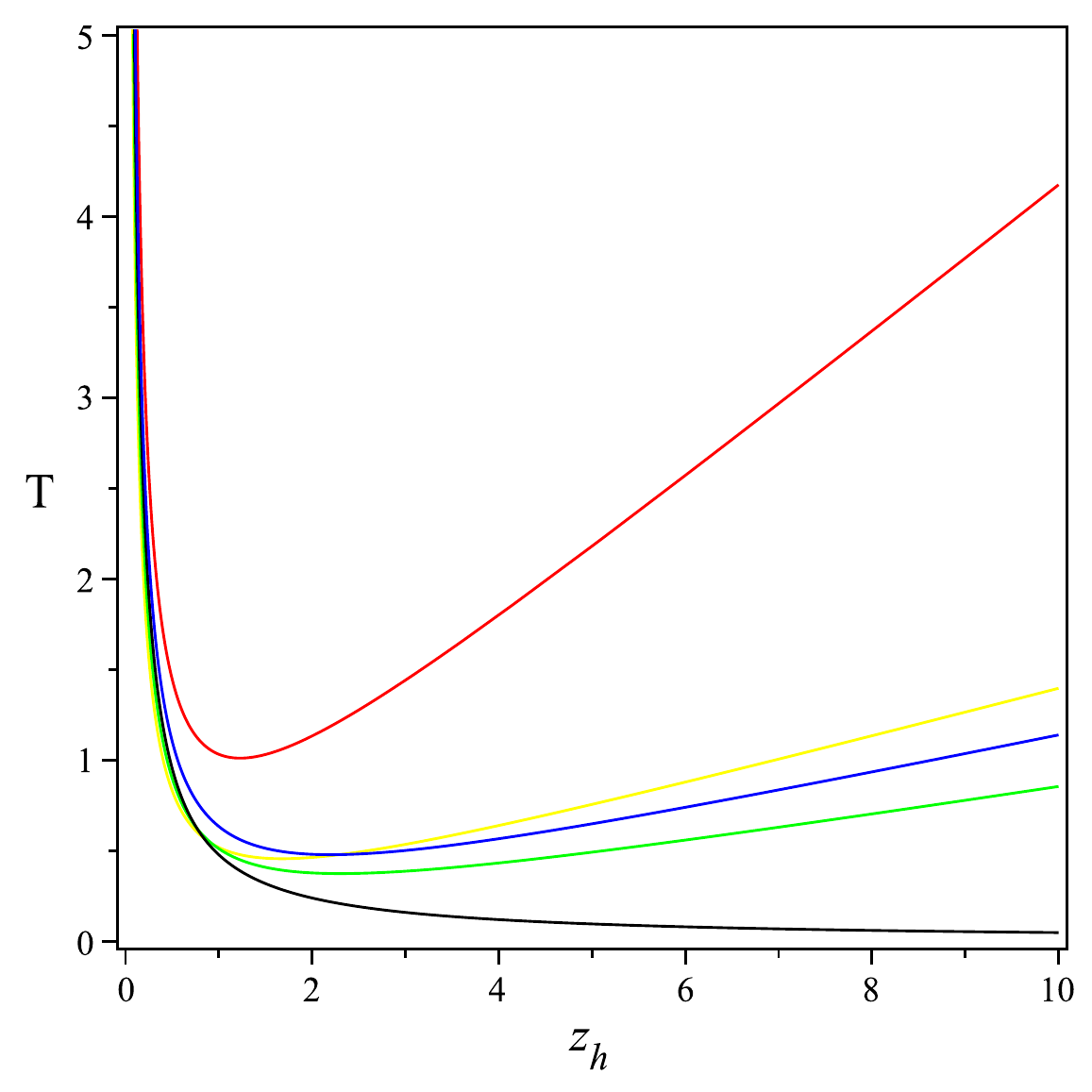}}
\subfigure{\includegraphics[width=0.3\columnwidth]{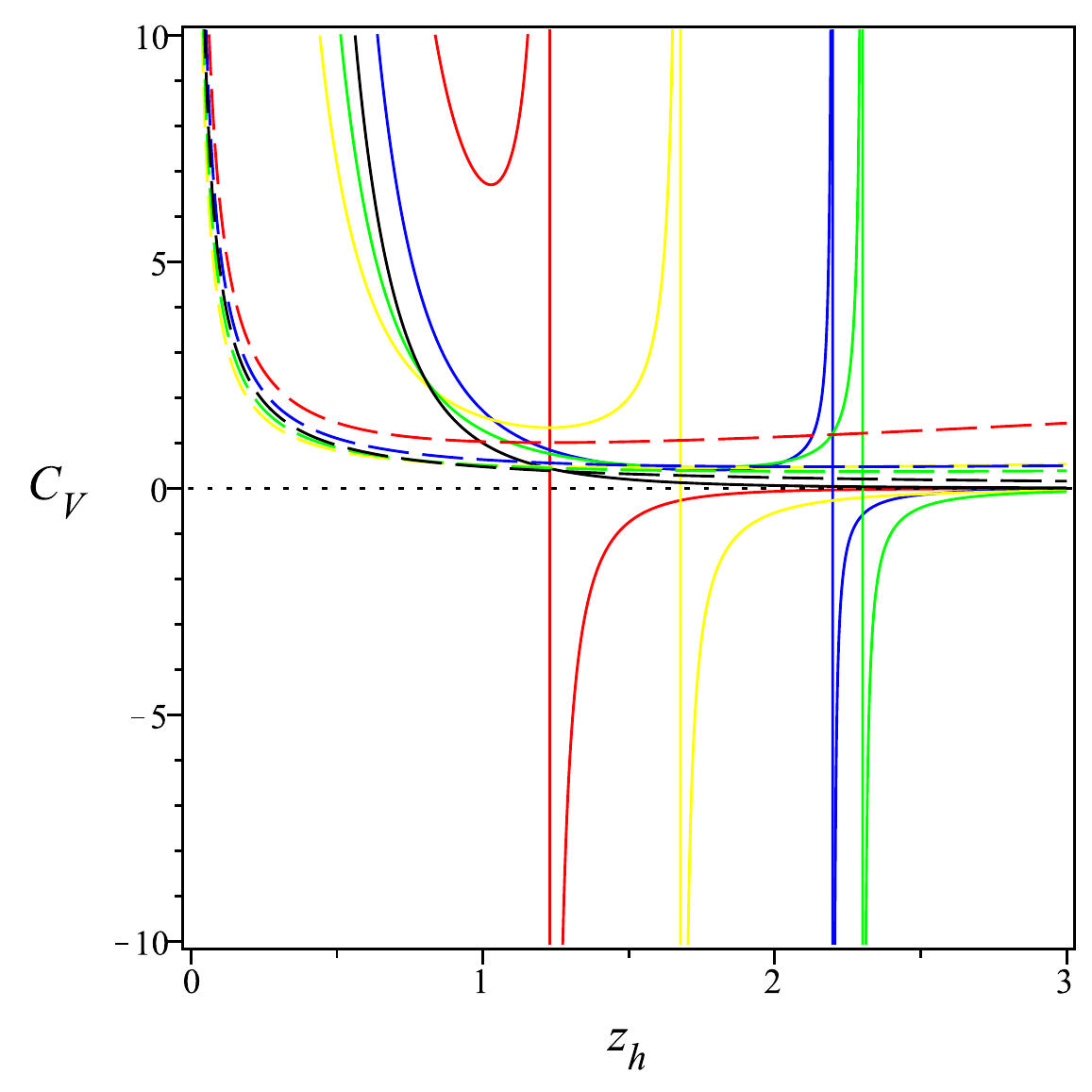}}
\subfigure{\includegraphics[width=0.3\columnwidth]{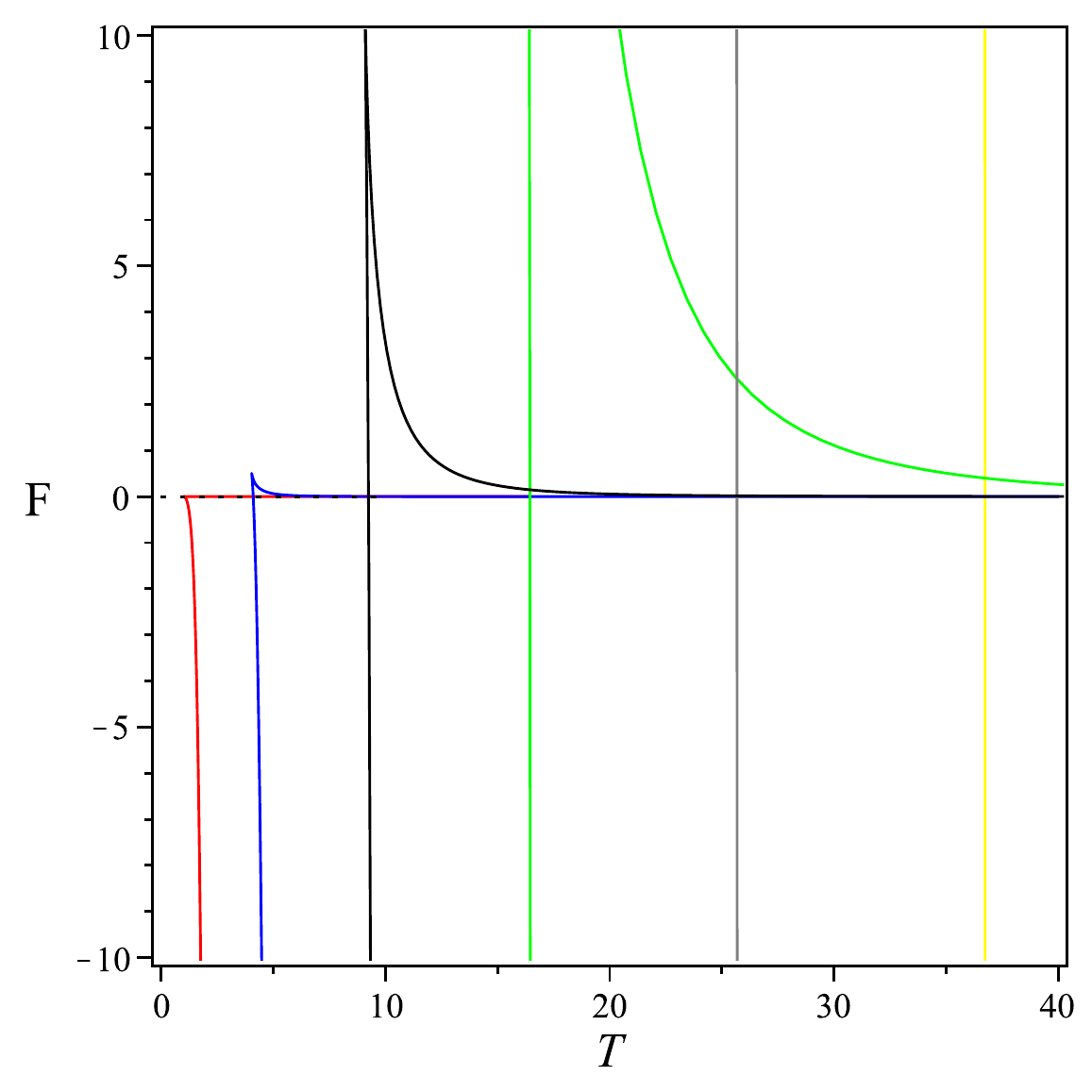}}
\caption{Plot of $T$ in terms of $z_{h}$ for  $k_{1}=-2,c=1,k_{0}=\textcolor{red}{1},\textcolor{blue}{-1},-2,\textcolor{green}{-3}$ (left). Plot of $C_{V}$ and $T$ in terms of $z_{h}$ for $c=1,2,3,4$ (middle). Plot of $F$ in terms of $T$ for different values of the anisotropy parameter $c=\textcolor{red}{1},\textcolor{blue}{2},3$ (right).} 
\label{Tzhplote}
\end{figure}
In Fig. (\ref{Tzhplote})b, to study the stability of the solutions, we have shown the behavior of heat capacity $C_{V}$ and temperature. As can be seen the large black hole has positive heat capacity and therefore is stable and small black hole is unstable and thus not physical.\\
Following the Wald formula, one can easily read the black hole
entropy density $S$, which is defined as
\begin{align}
S=-\dfrac{1}{8}\int_{\Sigma}d^{n-2}x\sqrt{\eta}\dfrac{\delta L}{\delta R_{\mu \alpha \beta \nu}}\epsilon_{\mu \alpha}\epsilon_{\beta \nu},
\end{align}
where
\begin{align}
\dfrac{\delta L}{\delta R_{\mu \alpha \beta \nu}}=&\left(\dfrac{1}{2}+\alpha R\right)\left(g^
{\mu \beta}g^{\alpha \nu}-g^{\mu \nu}g^{\alpha \beta}\right)+\dfrac{1}{2}\beta\left(R^{\mu \beta}g^{\alpha \nu}-R^{\alpha \beta}g^{\mu \nu}-R^{\mu \nu}g^{\alpha \beta}+R^{\alpha \nu}g^{\mu \beta}\right)\nonumber\\
&+2\gamma R^{\mu \alpha \beta \nu},
\end{align}
and $\epsilon_{\mu \nu}=-2\sqrt{-\zeta}\delta^{t}_{[\mu}\delta^{z}_{\nu]}$.
Here 
\begin{equation}
\sqrt{-\zeta}=A^{2},\;\;\;\;\sqrt{-\eta}=A^{3}h
\end{equation}
at the horizon the entropy becomes
\begin{align}\label{eqqent}
s=\left. \dfrac{1}{4}hA^{3}\right\vert_{z=z_{h}}- \left.\dfrac{\alpha g[3hA^{\prime 2}+2Ah^{\prime}A^{\prime}]}{4A}\right\vert_{z=z_{h}}=\left.\dfrac{1}{4}hA^{3}\right\vert_{z=z_{h}}=\dfrac{\sqrt{\mathfrak{c}}}{4}(cz_{h})^{\frac{3k_{1}-k_{0}-4}{2}}.
\end{align}
The scaled entropy density as a function of scaled temperature is shown in Fig. (\ref{szhplote})a. The red lines correspond to the large stable solution and blue lines are for the small unstable
solution \cite{Li:2011hp}.
The free energy density $F$ can be calculated from the entropy density $s$ by integrating as follows
\begin{small}
\begin{align}
F=&\int -sdT=\dfrac{(cz_{h})^{-\frac{1}{2}(k_{0}+4-3k_{1})}}{16\pi z_{h}}+\dfrac{\alpha \sqrt{\mathfrak{c}}(cz_{h})^{\frac{1}{2}(3k_{1}-k_{0}-6)}}{16\pi z_{h}c^2}+\dfrac{\alpha \sqrt{\mathfrak{c}}(k_{0}-3k_{1}+6)^{2}(k_{1}+1)(cz_{h})^{\frac{1}{2}(k_{1}-k_{0}-6)}}{64\pi(k_{0}-k_{1}+6)(k_{0}-4k_{1}+6)(k_{0}+6-4k_{1})}\nonumber\\
&[11k_{0}^{3}+(68-65k_{1})k_{0}^{2}+(131k_{1}^{2}-188k_{1}+12)k_{0}+60k_{1}
   +92k_{1}^{2}-81k_{1}^{3}]+\mathcal{O}(\alpha^2).
\end{align}
\end{small}
The $y_{1}y_{2}$-component of the pressures is related to the free energy as follows: $P_{y_{1}y_{2}}=-F$. 
{The other components of pressure can be obtained by an asymptotic expansion in the near boundary regime and reading the expectation value of the energy-momentum tensor components, by this expansion \cite{Mateos:2011tv},\cite{Mateos:2011ix}.}
The energy is given as follows:
\begin{small}
\begin{align}\label{eqenergy}
   E=&F+Ts=\dfrac{(k_{0}-3k_{1}+4)(cz_{h})^{\frac{3k_{1}-k_{0}-4}{2}}\sqrt{\mathfrak{c}}}{32\pi z_{h}}+{\dfrac{(k_{0}-3k_{1}+4)\sqrt{\mathfrak{c}}\alpha}{128\pi z_{h}^{3}c^{2}(k_{0}-k_{1}+6)(k_{0}-4k_{1}+6)(k_{0}-5k_{1}+6)}}\nonumber\\
   &[-4(k_{0}-k_{1}+6)(k_{0}-4k_{1}+6)(k_{0}+6-5k_{1})(cz_{h})^{\frac{3k_{1}-k_{0}}{2}}+c^{2}(cz_{h})^{\frac{k_{1}-k_{0}}{2}}(k_{0}+6-3k_{1})^{2}[11k_{0}^{3}+\nonumber\\
   &(68-65k_{1})k_{0}^{2}+(131k_{1}^{2}-188k_{1}+12)k_{0}+60k_{1}
   +92k_{1}^{2}-81k_{1}^{3}]]+\mathcal{O}(\alpha^2).
\end{align}
\end{small}
The unstable-stable nature of the small-large black hole phases can also be seen from the free energy behaviour shown in Fig. \ref{szhplote}. We see that the free energy of the small black hole phase is always larger than the large black hole and thermal AdS phases, indicating the unstable nature of this small black hole phase. Importantly, upon varying the Hawking temperature, a phase transition from the large black hole phase to thermal AdS phase takes place at a critical temperature $T_{c}$. This is the famous black hole-thermal AdS Page-Hawking phase transition.\\
The dependence of $T_{c}$ on $\alpha$ for different values of $k_{0}$ has been shown in Fig. \ref{Tzhplote1}. {There exists a naked singularity of the zero temperature solution.}

\begin{figure}[H]\hspace{0.4cm}
\centering
\subfigure{\includegraphics[width=0.45\columnwidth]{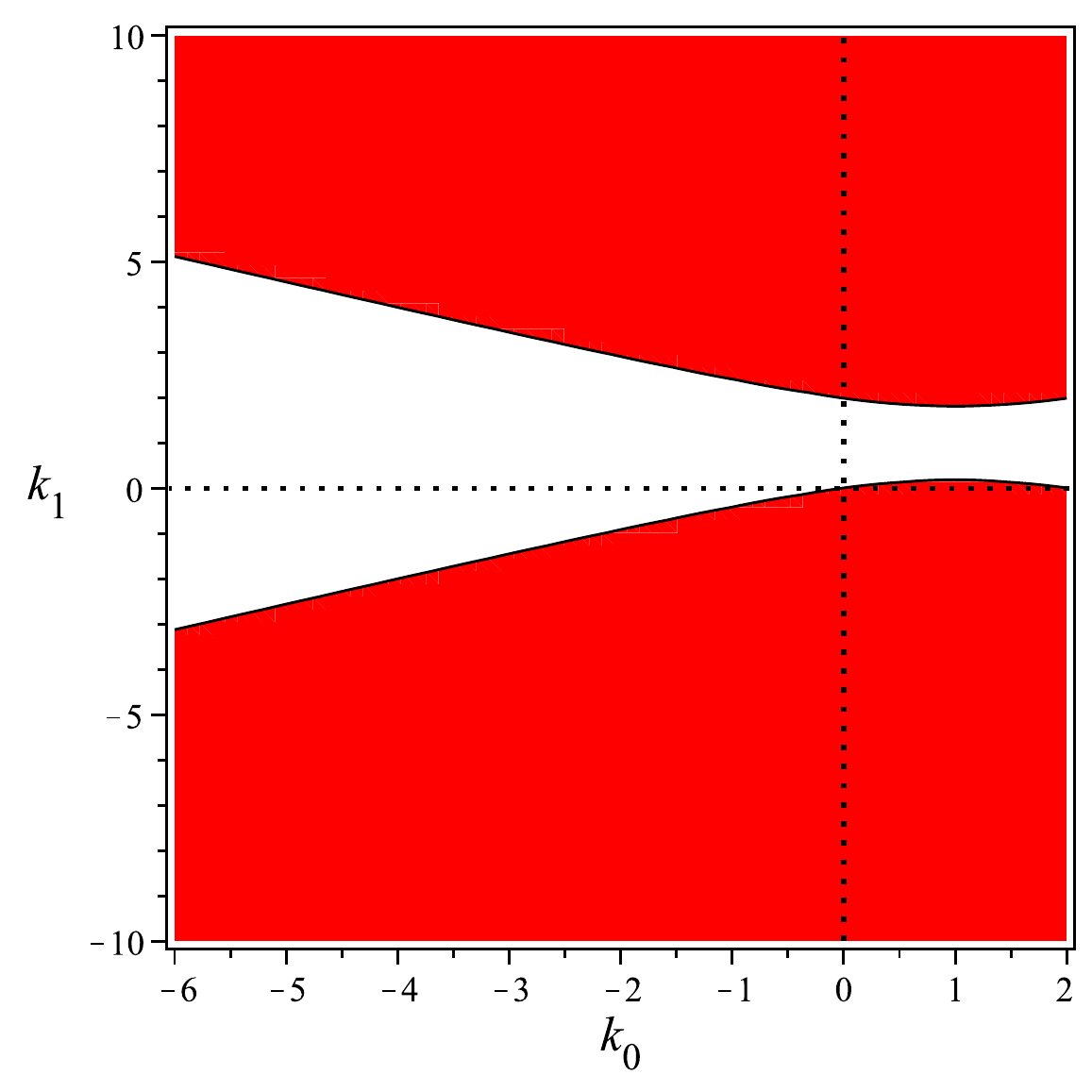}}
\caption{The allowed region for $k_{1}$ and $k_{0}$ are shown in red color.} 
\label{kk10plot}
\end{figure}

The sound velocity $c_{s}^{2}$ which can directly measure the conformality of the system, can be obtained from the temperature and entropy:
\begin{align}\label{vsound}
c_{s}^{2}=&\dfrac{d\log T}{d\log s}=\dfrac{2}{k_{0}+4-3k_{1}}-{\dfrac{\alpha k_{1}(k_{0}+6-3k_{1})c^{-k_{1}+2}z_{h}^{-k_{1}}}{2(k_{0}+6-5k_{1})(k_{0}+6-4k_{1})(k_{0}+4-3k_{1})}}\times\nonumber\\
&[11k_{0}^{3}+(68-35k_{1})k_{0}^{2}+(131k_{1}^{2}-188k_{1}+12)k_{0}-81k_{1}^{3}
+92k_{1}^{2}+60k_{1}]+\mathcal{O}(\alpha^{2}).
\end{align}
For isotropic case ($k_{0}+4-3k_{1}=6$), $c_{s}^{2}= 1/3$, the system is conformal, for anisotropic ($k_{0}+4-3k_{1}\neq6$), $c_{s}^{2}\neq 1/3$ the system is non-conformal. The numerical result of the square of the sound velocity is shown in \ref{szhplote}b. At $T_{min}$, the sound velocity square is around $0$ which is in agreement with lattice data $0.05$. At high temperature, the sound velocity square goes to $0.33$ for $k_{0}=-4$, which means that the system is asymptotically conformal. The heat capacity is given as
\begin{small}
\begin{align}
C_{V}=&T\dfrac{ds}{dT}=\dfrac{(k_{0}+4-3k_{1})\sqrt{\mathfrak{c}}c^{2}(cz_{h})^{\frac{3k_{1}-k_{0}-4}{2}}}{8}+{\dfrac{\alpha \sqrt{\mathfrak{c}} k_{1}(k_{0}+4-3k_{1})(k_{0}-3k_{1}+6)(cz_{h})^{\frac{-k_{0}-4+k_{1}}{2}}}{32(k_{0}-4k_{1}+6)(k_{0}-5k_{1}+6)}}\nonumber\\
&\times[11k_{0}^{3}+(68-35k_{1})k_{0}^{2}+(131k_{1}^{2}-188k_{1}+12)k_{0}-81k_{1}^{3}
+92k_{1}^{2}+60k_{1}]+\mathcal{O}(\alpha^{2})=\dfrac{s}{c_{s}^{2}}.
\end{align}
\end{small}
The numerical result of the specific heat is shown in Fig. \eqref{szhplote}c. It can be clearly seen that the specific heat $C_{V}$ diverges at $T_{min}$. At $T\to \infty$, the scaled specific heat $C_{V}/T^{3}$ approaches to the zero. {For a null vector $\xi^{\mu}$ the null energy condition is given as
\begin{align}\label{eqqcon}
T_{\mu \nu}\xi^{\mu}\xi^{\nu}\geq 0,
\end{align}
where $T_{\mu \nu}$ is the energy-momentum tensor which is given by the right-hand side of 
equation (\ref{eqq3}).
The magnitude of the null vector is zero. Thus
\begin{equation}
g_{\mu \nu}\xi^{\mu}\xi^{\nu}=0,
\end{equation}  
and using the geodesic equation leads to 
\begin{equation}
\xi^{\mu}=\dfrac{1}{A}\left[\dfrac{1}{g},\sqrt{1-\dfrac{g}{h}-2g},\dfrac{1}{h},1,1\right],
\end{equation}
using \eqref{eqqcon}, considering the leading terms of expansion, from the positivity of heat capacity, and boundness of the sound/butterfly velocities, one can obtain the conditions on $k_{0}$ and $k_{1}$ as follows:
\begin{align}
(2-k_{0})(k_{0}-3k_{1}+6)\geq 0,\\
3k_{1}^{2}-6k_{1}+2k_{0}-k_{0}^{2}\geq 0,\\
k_{0}-3k_{1}+4\geq 0,\\
k_{0}+6-5k_{1}\geq 0.
\end{align} 
Combining these inequalities, we observe that for $-6\leq k_{0}\leq2$ the value of $k_{1}$ is bounded from above $k_{1}^{(-)}$, and from below $k_{1}^{(+)}$ as
\begin{equation}
k_{1}^{(\pm)}=1\pm\dfrac{\sqrt{9-6k_{0}+3k_{0}^{2}}}{3}.
\end{equation}
The ranges $k_{0}>2$ and $k_{0}<-6$ are forbidden. The allowed range of parameters has been shown in figure \ref{kk10plot}.}

\subsection{Transport and Diffusion}\label{sec4}
Here, we look at the macroscopic and the microscopic motion of QGP using holography.
Transport is related to the movement of particles in response to an external force. 
Transport in the QGP is characterized by its viscosity, which is a measure of the resistance of the plasma to flow. The shear viscosity to entropy density ratio ($\eta/s$) is a quantity of particular interest, as it is a fundamental property that governs the evolution of the plasma. The holographic value of the shear viscosity $4\pi \eta/s=1$. 
This universal value is violated in anisotropic systems for the shear component parallel to the anisotropic direction, $\eta_{\vert\vert}$, while the component transverse to the anisotropic direction $\eta_{\bot\bot}$ remains universal. A calculation shows that $\eta_{\vert\vert}$ can be obtained in the IR limit
\begin{align}\label{eqqeta1}
\dfrac{\eta_{\parallel}}{s}=\dfrac{1}{4\pi}\dfrac{g_{\bot\bot}}{g_{\|}}=\dfrac{(cz_{h})^{k_{0}-2}}{\mathfrak{c}}.
\end{align}
The shear viscosity is parametrized according to the properties of the IR geometry, i.e. the
scaling exponents $k_{0}$ and $k_{1}$.
For the case of $k_{1}=0$ and $k_{0}=2-\epsilon$, equation \eqref{eqqeta1} becomes
\begin{equation}
\dfrac{4\pi \eta_{\parallel}}{s}=1-\ln\left(\dfrac{c}{\pi T}\right)\epsilon -\alpha \epsilon +\mathcal{O}(\alpha^2\epsilon^2),\;\;\;\epsilon\ll 1,\;\;\;\alpha\ll 1,
\end{equation}
where $\epsilon$ is deviation from the isotropy.
 We observe that the shear viscosity in the anisotropic direction is generally below the universal value $\eta/s=1/4\pi$, which is attained only in the UV regime.
Also, the GB term decreases the shear viscosity.
Another interesting phenomenon is momentum diffusion, which is related to the random motion of particles due to fluctuations in the local environment.
 In holographic theories, diffusion is characterized by the time scale $\tau=1/T$ and the butterfly velocity $v_{B}$, both entering in the diffusion constant as $D\sim v_{B}^{2}/T$. These quantities can be computed holographically through the near horizon
dynamics. In anisotropic theories, there are two notions of butterfly velocities, $v_{B\parallel}$ and $v_{B\perp}$, corresponding to the parallel and transverse directions, respectively. On the background one finds
\begin{align}
\mu_{\perp}^{2}&=\left. \dfrac{g^{\prime}(3k_{1}-k_{0}-4)}{4z}\right\vert_{z=z_{h}}=\dfrac{(k_{0}+4-3k_{1})(k_{0}+6-3k_{1})}{8z_{h}^{2}}-\dfrac{(k_{0}+6-3k_{1})(k_{0}+4-3k_{1})\alpha}{32z_{h}^{2}(k_{0}+6-5k_{1})(k_{0}+6-4k_{1})}\nonumber\\
&[(k_{0}+6-3k_{1})c^{2-k_{1}}(k_{0}^3+k_{0}^{2}(68-65k_{1})+k_{0}(12+131k_{1}^{2}-188k_{1})+92k_{1}^{2}+60k_{1}-81k_{1}^{3})-\nonumber\\
&4(k_{0}+6-4k_{1})(k_{0}+6-5k_{1})]+\mathcal{O}(\alpha^{2}),\\
\mu_{\parallel}^{2}&=\dfrac{\mathfrak{c}\mu_{\perp}^{2}}{(cz_{h})^{k0-2}}.
\end{align}
The corresponding butterfly velocities $v_{Bi}^{2}=(2\pi T)^{2}/\mu_{i}^{2}$ in the IR regime becomes
\begin{align}
v_{B\perp}^{2}&=\dfrac{k_{0}+6-3k_{1}}{2k_{0}+8-6k_{1}}+\dfrac{\alpha(k_{0}+6-3k_{1})}{8(k_{0}+6-4k_{1})(k_{0}+6-5k_{1})(k_{0}+4-3k_{1})}[(k_{0}+6-3k_{1})z_{h}^{-k_{1}}c^{-k_{1}+2}\nonumber\\
&(11k_{0}^{3}+k_{0}^{2}(76-69k_{1})+k_{0}(60-244k_{1}+147k_{1}^{2})-93k_{1}^{3}+12k_{1}+140k_{1}^{2})+4(k_{0}+6-5k_{1})\nonumber\\
&(k_{0}+6-4k_{1})]+\mathcal{O}(\alpha^2),\\
v_{B\parallel}^{2}&=\dfrac{(cz_{h})^{k_{0}-2}}{\mathfrak{c}}v_{B\perp}^{2}.
\end{align}
For $k_{0}=2,k_{1}=0$, the velocities becomes $2/3+\alpha$. $v_{B\perp}^{2}\to 3/4$ as $k_{0}=k_{1}=0$. 
For other values $k_{0}=2-\epsilon$ and $k_{1}=0$, we have
\begin{equation}
\dfrac{3}{2}v_{B\perp}^{2}=1+\alpha+\epsilon\left[0.04-(1+\alpha)\ln\left(\dfrac{c}{\pi T}\right)-\alpha\right]+\mathcal{O}(\alpha^{2}\epsilon^{2}).\\
\end{equation}
For allowed values of parameters,  $v_{B\parallel}^{2}$ and $v_{B\perp}^{2}$ are positive and  $v_{B\parallel}^{2}$ smaller than 2/3 but the value of $v_{B\perp}^{2}$ in the IR regime can exceed the conformal value 2/3. For conformal but anisotropic case in GB theory provided in appendix \ref{app1}. 

\subsection{The Imaginary Part of the Potential}\label{sec5}
Here, we study the imaginary part of the static potential associated to the thermal width in finite temperature strongly coupled anisotropic plasma \cite{BitaghsirFadafan:2013vrf},\cite{Ali-Akbari:2014vpa},\cite{Finazzo:2013rqy}.
The imaginary part of the quark-antiquark potential arises due to the presence of virtual quark-antiquark pairs, which can be created and annihilated in the interaction between two heavy quarks, which can be extracted from the correlation functions of heavy quark-antiquark pairs.
The real part of the static potential is related to the energy of the system, while the imaginary part is related to the decay rate of the system. The imaginary part of potential is given by
\begin{equation}
    Im V_{Q\bar{Q}}=-\dfrac{1}{2\sqrt{2}\lambda}\sqrt{M_{0}}\left[\dfrac{V^{\prime}_{0}}{V^{\prime\prime}_{0}}-\dfrac{V_{0}}{V^{\prime}_{0}}\right]
\end{equation}
where
\begin{align}
    M=&-g_{tt}g_{yy}=A^{4}(z)g(z)(\sin^{2}{\theta}+h^{2}(z)\cos^{2}{\theta}),\nonumber\\
    V=&-g_{tt}g_{zz}=A^{4}(z),
\end{align}
and $V_{0}=V(z_{0})$, and $z_{0}$ is the minimum of string. For generic values of $k_{0}$, $k_{1}$ and $\theta$, the imaginary part of potential has a huge, complicated expression. Therefore, we consider the special cases:
for $\theta=\pi/2$ and $k_{0}=8$ and $k_{1}=2+\epsilon$ (conformal-anisotropic) becomes
\begin{align}
    Im V_{Q\bar{Q}}\sim &-\dfrac{\sqrt{2}(-1+3Z_{0}^4)}{48\pi \lambda T Z_{0}}+\dfrac{\sqrt{2}\epsilon}{576\pi\lambda Z_{0}T}[12(3Z_{0}^{4}-1)\ln\left(\dfrac{\pi T}{c}\right)+6(15Z_{0}^4-2)\ln Z_{0}+\nonumber\\
    &(52Z_{0}^4-18Z_{0}^{8}-5)]-\dfrac{\sqrt{2}(-1+3Z_{0}^2)\alpha}{48\pi \lambda T Z_{0}}+\mathcal{O}(\alpha\epsilon).
\end{align}
For $\theta=0$, we have 
\begin{align}
  Im V_{Q\bar{Q}}&\sim  -\dfrac{(1+\alpha)\sqrt{2}(6Z_{0}^{8}-9Z_{0}^{4}+1)}{12\pi\lambda Z_{0}T(-1+3Z_{0}^{4})(-1+7Z_{0}^{4})} -\dfrac{21\epsilon}{\sqrt{2}\pi Z_{0}\lambda T(1-7Z_{0}^{4})^{2}(1-3Z_{0}^{2})^{2}}[-(\frac{83}{42}Z_{0}^{12}+\nonumber\\
  &\frac{19}{126}Z_{0}^{4}-Z_{0}^{16}-Z_{0}^{8}-\frac{1}{126})\ln(\frac{\pi T}{c})-\frac{17Z_{0}^{4}-1}{756}+\frac{41Z_{0}^{16}}{168}-\frac{17Z_{0}^{12}}{42}+\frac{11Z_{0}^{8}}{72}] +\mathcal{O}(\epsilon^{2}\alpha^{2}).
\end{align}
For $k_{1}=0$, $k_{0}=2+\epsilon$ and $\theta=0,\pi/2$ we have
\begin{align}
 Im V_{Q\bar{Q}}&\sim -\dfrac{\sqrt{2}\pi T(-5+3Z_{0}^{4})}{80\lambda c^{2}Z_{0}^{3}}-\dfrac{\sqrt{2}\alpha\pi T(-15Z_{0}^{12}+30c^{2}Z_{0}^{12}+25Z_{0}^{8}-100c^{2}Z_{0}^{8}+132c^{2}Z_{0}^{4}-50c^{2})}{400\lambda c^{2}Z_{0}^{11}}\nonumber\\
 &+\dfrac{\sqrt{2}\pi\epsilon T(25-8Z_{0}^{4}-15Z_{0}^{8}+100Z_{0}^{4}\ln(Z_{0}))}{3200\lambda c^{2}Z_{0}^{7}}+\mathcal{O}(\epsilon\alpha), 
\end{align}
where $Z_{0}=z_{h}/z_{0}$.

\subsection{Jet Quenching}\label{sec6}

Jet quenching is related to the suppression of high-energy jets of particles produced in high-energy heavy-ion collisions due to their interaction with QGP. The jet quenching parameter for a parton (quark or gluon) moving along the $p$ direction while the broadening happens along the $k$ direction is given by \cite{Giataganas:2012zy}:
\begin{align}
q_{p}(k)=\dfrac{\sqrt{2}}{\pi \alpha}\left(\int_{0}^{z_{h}}dz\dfrac{1}{g_{kk}}\sqrt{\dfrac{g_{zz}}{g_{--}}}\right)^{-1},\;\;\; g_{--}=\dfrac{1}{2}(g_{tt}+g_{pp}).
\end{align}
First assume the quark moves along the anisotropic direction and the momentum broadening occurs along the transverse one. Then
\begin{equation}
    g_{pp}=g_{xx},\;\;\;\;g_{kk}=g_{y_{1}y_{1}}
\end{equation}
for this case
\begin{equation}
    \dfrac{q_{\parallel}(\perp)}{q_{0}}\sim 1+\dfrac{\epsilon \sqrt{c_{1}}}{c^{3}}\left[-0.25+\ln\left(\dfrac{0.32c}{T}\right)\right]+0.5\dfrac{\sqrt{c_{1}\alpha}}{c^{3}}+\mathcal{O}(\epsilon\alpha).
\end{equation}
The second case is
\begin{equation}
    g_{pp}=g_{y_{1}y_{1}},\;\;\;\;g_{kk}=g_{xx},
\end{equation}
for this case $q_{\perp}(\parallel)$ we have
\begin{equation}
\dfrac{q_{\perp}(\parallel)}{q_{0}}\sim 1+\dfrac{\epsilon {c_{1}}}{c^{6}}\left[-0.14+\ln\left(\dfrac{0.32c}{T}\right)\right]+0.75\dfrac{{c_{1}\alpha}}{c^{6}}+\mathcal{O}(\epsilon\alpha).
\end{equation}
Finally, we look at $q_{\perp}(\perp)$ where the quark motion and the momentum broadening happen along the transverse directions as follows
\begin{equation}
    g_{pp}=g_{y_{1}y_{1}},\;\;\;\;\;g_{kk}=g_{y_{2}y_{2}}.
\end{equation}
Therefore
\begin{equation}
   \dfrac{ q_{\perp}(\perp)}{q_{0}}\sim 1+\epsilon \left[-0.88+0.95\ln\left(\dfrac{0.32c}{T}\right)\right]+0.7\alpha +\mathcal{O}(\epsilon\alpha).
\end{equation}
To summarize we find that the jet quenching is in generally enhanced in presence of anisotropy compared to the isotropic case and that its value depends on the direction of the moving quark and the direction which the momentum broadening occurs. But the effect of GB term is more respect to the anisotropy directions. More particularly
\begin{equation}
    q_{\perp}(\parallel)>q_{\parallel}(\perp)>q_{\perp}(\perp)>q_{0}
\end{equation}

\begin{figure}[H]\hspace{0.4cm}
\centering
\subfigure{\includegraphics[width=0.45\columnwidth]{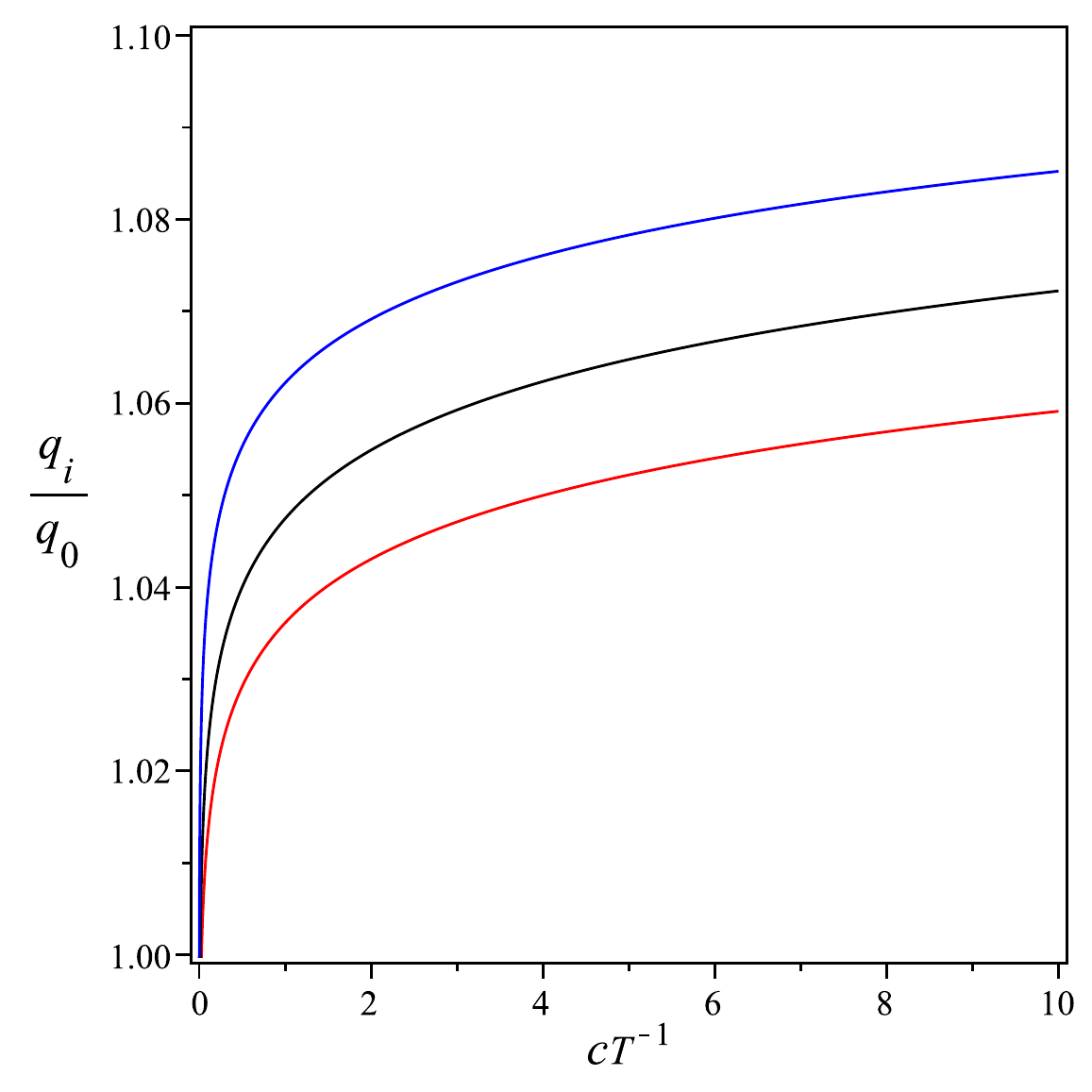}}
\caption{Plot of $\textcolor{red}{q_{\perp}(\parallel)/q_{0}}$, $\textcolor{black}{q_{\perp}(\perp)/q_{0}}$ and $\textcolor{blue}{q_{\parallel}(\perp)/q_{0}}$ in terms of $c/T$ for $\alpha=0.1,\epsilon=0.01$.} 
\label{Tzhplote}
\end{figure}

\section{Conclusion}\label{conclud}
In this work, we extended the strongly coupled anisotropic systems at finite temperature to quadratic gravity to gain insight into the influence of gravity on QCD. To do so, we considered an anisotropic black hole metric as a solution to a system of $5D$ Quadratic-Axion-Dilaton Gravity. The anisotropic background is specified by an arbitrary exponent ($k_{0}$ and $k_{1}$), a non-zero dilaton field, a non-zero axion field. The field equations for the considered theory are coupled and bulky differential equations for six unknown functions. Therefore, obtaining the solution to such field equations is too hard, this is why we considered the special case of field equation, i.e. $\gamma=\alpha, \beta=-4\alpha$ (Einstien-Gauss-Bonnet gravity). The differential equation for the metric function in EGB gravity is a nonlinear second-order equation that has been solved in a perturbative manner. In addition to the small/large phase transition, the blackening function supported the phase transition from the large black hole phase to the thermal AdS phase at a critical temperature. Holographically, this phase transition corresponds to the confinement-deconfinement phase transition in QCD. In each case, we investigated the anisotropy influence and the effect of parameters of theory on the thermodynamic properties of our background, in particular, on the small/large black holes phase transition diagram.
Finally, we study transport and diffusion properties in anisotropic theories and observe in particular that the butterfly velocity that characterizes both diffusion and growth of chaos transverse to the anisotropic direction saturates a constant value in the IR limit which can exceed the bound given by the conformal value.\\
For future work, one can look at the effect of GB term on the entropic $c$-function following \cite{Chu:2019uoh} the pressure along the different directions, the velocity law \cite{Giataganas:2022wqj}, bound states in baryons \cite{Giataganas:2018uuw} and on the rotation of the mesons \cite{Giantsos:2022qgq}.\\

{\bf Acknowledgements}\\
We would like to thank Dimitrios Giataganas for his fruitful comments which help us to improve the manuscript. We also thank the anonymous referee for her/his valuable comments.

\appendix

\section*{The result about conformal-anisotropic solution}\label{app1}
From the equation \eqref{temp}, to have a result of Schwartzschild-AdS$_{5}$ we should have
\begin{align}
  k_{0}-3k_{1}-2=0.
  \end{align}
  To obtain the results of conformal and anisotropic, we write 
  \begin{equation}\label{confcond}
  k_{0}=8,\;\;\;\; k_{1}=2+\epsilon \;\;\;\;\; \epsilon\ll 1
  \end{equation}
 for which $\epsilon$ is the deviation from the isotropicity. Under the conditions \eqref{confcond}, we presented some results below:

\begin{equation}
   T=\dfrac{1+\alpha}{\pi z_{h}}+\epsilon\left[-\dfrac{3}{8\pi z_{h}}-\alpha\left(-\dfrac{3}{8\pi z_{h}}+\dfrac{8}{\pi z_{h}^{3}}\right)\right]+\mathcal{O}(\alpha^2,\epsilon^2).
\end{equation}

\begin{align}
    s\sim &\dfrac{\sqrt{\mathfrak{c}}\pi ^{3}T^{3}}{4c^{3}}-\dfrac{3\sqrt{\mathfrak{c}}\alpha \pi^{3}T^{3}}{4c^{3}}+\dfrac{\epsilon\sqrt{\mathfrak{c}}\pi^{3}T^{3}}{8c^{3}}\left[\dfrac{9}{4}+\ln\left(\dfrac{c}{\pi T}\right)\right]+\nonumber\\
    &\dfrac{\epsilon\alpha\sqrt{\mathfrak{c}}\pi^{3}T^{3}}{8c^{3}}\left[3\ln\left(\dfrac{\pi T}{c}\right)-\dfrac{23}{4}\right]+\mathcal{O}(\alpha^2,\epsilon^{2}).
\end{align}

\begin{align}
    E=&\dfrac{3\pi^{3}\sqrt{\mathfrak{c}}T^{4}}{16c^{3}}-\dfrac{9\alpha \pi^{3}\sqrt{\mathfrak{c}}T^{4}}{16c^{3}}-\dfrac{3\epsilon\pi^{3}\sqrt{\mathfrak{c}}T^{4}}{32c^{3}}\left[\ln\left(\dfrac{\pi T}{c}\right)-\dfrac{13}{2}\right]-\nonumber\\
    &\dfrac{9\alpha\epsilon\pi^{3}\sqrt{\mathfrak{c}}T^{4}}{32c^{3}}\left[\ln\left(\dfrac{c}{\pi T}\right)+\dfrac{11}{6}\right]+\mathcal{O}(\alpha^2,\epsilon^{2}).
\end{align}

\begin{align}\label{eqqeta}
\dfrac{\eta_{\vert\vert}}{s}\sim\dfrac{c^{6}}{\mathfrak{c}\pi^6T^{6}}+\dfrac{6\alpha c^{6}}{\mathfrak{c}\pi^6T^{6}}-\dfrac{9\epsilon c^{6}}{4\mathfrak{c}\pi^{6}T^{6}}+\mathcal{O}(\alpha^2,\epsilon^2)
\end{align}

\begin{align}
\mu_{\perp}^{2}&\sim 6\pi^2 T^{2}-6\alpha \pi^2 T^{2}(1+32\pi^2 T^{2})-\dfrac{3\epsilon \pi^{2}T^{2}}{4}+\mathcal{O}(\alpha\epsilon)\nonumber\\
\mu_{\parallel}^{2}&\sim \dfrac{6\mathfrak{c}\pi^{8}T^{8}}{c^{6}}+\dfrac{51\epsilon\mathfrak{c}\pi^{8}T^{8}}{4c^{6}}-\dfrac{6\mathfrak{c}\pi^{8}T^{8}}{c^{6}}(7+232\pi^{2}T^{2})+\mathcal{O}(\epsilon\alpha).
\end{align}

\begin{align}
 v_{B\perp}^{2}&\sim\dfrac{2}{3}+\dfrac{\epsilon}{12}+\alpha\Big[\dfrac{3}{2}+\dfrac{464\pi^2 T^{2}}{3}\Big]+\mathcal{O}(\alpha\epsilon),\nonumber\\
 v_{B\parallel}^{2}&\sim\dfrac{2c^{6}}{3\mathfrak{c}\pi^{6}T^{6}}+\alpha\Big[\dfrac{14c^{6}}{3\mathfrak{c}\pi^{6}T^{6}}+\dfrac{464c^{6}}{3\mathfrak{c}\pi^{4}T^{4}}\Big]-\dfrac{17\epsilon c^{6}}{12\mathfrak{c}\pi^{6}T^{6}}+\mathcal{O}(\alpha\epsilon),
\end{align}
where we have used
\begin{equation}
   z_{h}\sim\dfrac{1+\alpha}{\pi T}-\dfrac{3\epsilon}{8\pi T}-\dfrac{3\epsilon\alpha}{8\pi T}+\mathcal{O}(\alpha^2,\epsilon^{2}).
\end{equation}
We see that in the limit $\epsilon\to 0$ the results approach Schwartzschild-AdS$_{5}$, as expected.


\section*{The components of the field equation}\label{app2}
Here, the components of the field equation are presented:
\begin{eqnarray}
&&E_{zz}=-6A^2hA^{\prime\prime}-2A^{3}h^{\prime\prime}+12AhA^{\prime 2}-hA^{3}\phi^{\prime 2}+\dfrac{\alpha}{A^{3}}\left[{24hgAA^{\prime 2}A^{\prime\prime}}+{16gA^{2}h^{\prime}A^{\prime}A^{\prime\prime}}-{48ghA^{\prime 4}}\right.\nonumber\\
&&\left. \;\;\;\;\;\;-{32gAh^{\prime}A^{\prime 3}}+{8gA^{2}A^{\prime 2}h^{\prime\prime}}\right]=0,\\
&&E_{tt}=-48AhgA^{\prime\prime}-6hA^{2}g^{\prime\prime}-12A^2gh^{\prime\prime}-
48AhA^{\prime}g^{\prime}-48gAA^{\prime}h^{\prime}-24ghA^{\prime 2}+10hA^{4}V(\phi)\nonumber\\
&&\;\;\;\;-12A^2g^{\prime}h^{\prime}-3hgA^{2}\phi^{\prime 2}-\dfrac{3c_{1}^2A^2Z(\phi)}{h}+\dfrac{\alpha}{A^4}\left[192Ahg^2A^{\prime\prime}A^{\prime 2}+96g^2A^2A^{\prime}A^{\prime\prime}h^{\prime}+48hgA^{2}
g^{\prime}A^{\prime}A^{\prime\prime}\right.\nonumber\\
&&\left. \;\;\;\;+16A^{3}gA^{\prime\prime}h^{\prime}g^{\prime}+24hgA^{2}g^{\prime\prime}A^{\prime 2}+16gA^{3}A^{\prime}h^{\prime}g^{\prime\prime}+
16gA^3A^{\prime}g^{\prime}h^{\prime\prime}+48g^2A^2h^{\prime\prime}A^{\prime 2}+96hAgA^{\prime 3}g^{\prime}\right.\nonumber\\
&&\left. \;\;\;\;+{112gA^2g^{\prime}h^{\prime}A^{\prime 2}}+24hA^2g^{\prime 2}A^{\prime 2}-144hg^{2}A^{\prime 4}+16A^{3}g^{\prime 2}A^{\prime}h^{\prime}\right]=0,\\
&&E_{xx}=-12A^{2}gA^{\prime\prime}-2A^{3}g^{\prime\prime}-
12A^{2}A^{\prime}g^{\prime}+2A^{5}V(\phi)-gA^{3}\phi^{\prime 2}+\dfrac{c_{1}A^{3}Z(\phi)}{h^{2}}+\dfrac{\alpha}{A^{3}}\left[16gA^{2}A^{\prime}g^{\prime}A^{\prime\prime}\right.\nonumber\\
&&\left. \;\;\;\;\;+48Ag^{2}A^{\prime 2}A^{\prime\prime}+16gA^{2}g^{\prime}A^{\prime 3}+8A^{2}A^{\prime 2}g^{\prime 2}+8gA^{2}A^{\prime 2}g^{\prime\prime}-48g^{2}A^{\prime 4}\right]=0,\\
&&E_{y_{1}y_{1}}=-6hgA^{2}A^{\prime\prime}-6hA^{2}A^{\prime}g^{\prime}-
6gA^{2}h^{\prime}A^{\prime}-hA^{3}g^{\prime\prime}-
\dfrac{c_{1}^{2}A^{3}Z(\phi)}{2h}-\dfrac{1}{2}hgA^{3}\phi^{\prime 2}+hA^{5}V(\phi)-\nonumber\\
&&\;\;\;\;\;\;\;2A^{3}g^{\prime}h^{\prime}-2A^{3}gh^{\prime\prime}+\dfrac{\alpha}{A^{3}}\left[24hAg^{2}A^{\prime 2}A^{\prime\prime}+8hA^{2}gA^{\prime}g^{\prime}A^{\prime\prime}+
16A^{2}g^{2}h^{\prime}A^{\prime}A^{\prime\prime}+
4gA^{3}h^{\prime}g^{\prime}A^{\prime\prime}-\right.\nonumber\\
&&\left. \;\;\;\;\;\;24hg^{2}A^{\prime 4}+4gA^{3}A^{\prime}h^{\prime}g^{\prime\prime}+4ghA^{2}A^{\prime 2}g^{\prime\prime}+4hA^{2}g^{\prime 2}A^{\prime 2}+16gA^{2}h^{\prime}g^{\prime}A^{\prime 2}+4gA^{3}A^{\prime}g^{\prime}h^{\prime\prime}+\right.\nonumber\\
&&\left. \;\;\;\;\;\;8g^{2}A^{2}A^{\prime 2}h^{\prime\prime}+4A^{3}A^{\prime}h^{\prime}g^{\prime 2}+8ghAg^{\prime}A^{\prime 3}-8Ag^{2}h^{\prime}A^{\prime 3}\right]=0.
\end{eqnarray}

\section*{The GB Correction to Potentials}\label{app3}
Here, we present the GB correction to equations \eqref{zphi} and \eqref{Vphi} as follows:
\begin{align}
\hat{Z}=&\dfrac{-(k_{0}-2)}{4c_{1}k_{1}z_{h}^{6}(k_{0}-4k_{1}+6)(k_{0}+6-5k_{1})(-3k_{1}^{2}+6k_{1}+k_{0}^{2}-2k_{0})}\Big[8k_{1}(k_{1}-2)(k_{0}+6-4k_{1})\nonumber\\
&z_{h}^{3+\frac{3k_{1}}{2}-\frac{k_{0}}{2}}c^{4-k_{0}-k_{1}}e^{\frac{(k_{0}+5k_{1}-6)\sqrt{6k_{1}^{2}-12k_{1}-2k_{0}^{2}+4k_{0}}(2\phi+\ln(z_{h})\sqrt{6k_{1}^{2}-12k_{1}-2k_{0}^{2}+4k_{0}})}{24k_{1}-12k_{1}^{2}+4k_{0}^{2}-8k_{0}}}\Big(-\frac{3k_{1}^{5}}{4}+(-\frac{3}{2}+\nonumber\\
&\frac{45k_{0}}{4})k_{1}^{4}+(-\frac{17k_{0}^{2}}{4}-41k_{0}+15)k_{1}^{3}+(37k_{0}-\frac{13}{4}k_{0}^{3}-18+21k_{0}^{2})k_{1}^{2}+k_{0}^{2}(4k_{0}-24+k_{0}^{2})k_{1}\nonumber\\
&+6k_{0}^{2}-\frac{k_{0}^{4}}{2}-2k_{0}^{3}\Big)+(k_{0}+6-5k_{1})(k_{0}+6-4k_{1})(k_{1}-2)^{2}z_{h}^{6}c^{4-k_{0}-k_{1}}\nonumber\\
&e^{\frac{(k_{0}+k_{1})\sqrt{6k_{1}^{2}-12k_{1}-2k_{0}^{2}+4k_{0}}(2\phi+\ln(z_{h})\sqrt{6k_{1}^{2}-12k_{1}-2k_{0}^{2}+4k_{0}})}{12k_{1}-6k_{1}^{2}+2k_{0}^{2}-4k_{0}}}(-6k_{1}^{4}+(48+15k_{0})k_{1}^{3}-(7k_{0}^{2}+40k_{0}+72)k_{1}^{2}\nonumber\\
&+
(12k_{0}^{2}-3k_{0}^{3}+36k_{0})k_{1}+k_{0}^{2}(k_{0}+6)(k_{0}-2))+(7k_{1}^{2}-(9k_{0}+10)k_{1}+2k_{0}^{2}+14k_{0}+12)(k_{0}+\nonumber\\
&6-5k_{1})^{2}(k_{1}-2)k_{1}(6k_{1}+k_{0}^{2}-2k_{0}-3k_{1}^{2})c^{4-k_{0}-k_{1}}z_{h}^{3k_{1}-k_{0}}\nonumber\\
&e^{\frac{(-3+2k_{1})\sqrt{6k_{1}^{2}-12k_{1}-2k_{0}^{2}+4k_{0}}(2\phi+\ln(z_{h})\sqrt{6k_{1}^{2}-12k_{1}-2k_{0}^{2}+4k_{0}})}{6k_{1}-3k_{1}^{2}+k_{0}^{2}-2k_{0}}}-(-4k_{1}+k_{0}+6)(k_{0}+6-3k_{1})\nonumber\\
&e^{\frac{k_{0}\sqrt{6k_{1}^{2}-12k_{1}-2k_{0}^{2}+4k_{0}}(2\phi+\ln(z_{h})\sqrt{6k_{1}^{2}-12k_{1}-2k_{0}^{2}+4k_{0}})}{12k_{1}-6k_{1}^{2}+2k_{0}^{2}-4k_{0}}}(k_{0}(k_{1}-2)^2(k_{0}+6-3k_{1})(k_{0}+3k_{1}-2)\nonumber\\
&(k_{0}-k_{1})z_{h}^{6-k_{1}}c^{4-k_{0}-k_{1}}-2k_{1}c^{-k_{0}+2}z_{h}^{6}(k_{0}+6-5k_{1})(6k_{1}-3k_{1}^{2}+k_{0}^{2}-2k_{0}))\Big],
\end{align}

\begin{align}
\hat{V}=&-\dfrac{c^2(k_{1}-2)e^{-\frac{(k_{0}-k_{1}+6)\sqrt{6k_{1}^2-12k_{1}-2k_{0}^{2}+4k_{0}}\phi}{k_{0}^{2}+6k_{1}-3k_{1}^{2}-2k_{0}}}}{8(k_{0}^2+6k_{1}-3k_{1}^2-2k_{0})(k_{0}-5k_{1}+6)(k_{0}-4k_{1}+6)z_{h}^{6}}\Big[4c^{-2k_{1}+2}z_{h}^{-2k_{1}+6}(k_{0}-4k_{1}+6)\nonumber\\
&e^{\dfrac{(k_{0}+5k_{1}+6)\sqrt{6k_{1}^{2}-12k_{1}-2k_{0}^{2}+4k_{0}}\phi}{2k_{0}^{2}+12k_{1}-6k_{1}^{2}-4k_{0}}}\Big(9k_{1}^{6}-(45+\frac{9}{2}k_{0})k_{1}^{5}+(\frac{21}{2}k_{0}^{2}-12k_{0}+90)k_{1}^{4}+\nonumber\\
&(-\frac{9}{2}k_{0}^{3}-33k_{0}^{2}-108+102k_{0})k_{1}^{3}+(35k_{0}^{3}-\frac{7}{2}k_{0}^{4}+72-104k_{0}^{2}-120k_{0})k_{1}^{2}+(80k_{0}^{2}+
k_{0}^{5}+\nonumber\\
&4k_{0}^{4}-52k_{0}^{3})k_{1}-k_{0}^{2}(k_{0}+6)(k_{0}-2)^{2}\Big)-3c^{-2k_{1}+2}z_{h}^{-2k_{1}+6}(k_{1}-2)^{2}(k_{0}+6-4k_{1})(k_{0}-k_{1}-\nonumber\\
&2)(k_{0}-5k_{1}+6)(k_{0}^{2}-2k_{0}+3k_{1}^{2}-6k_{1})e^{\frac{(k_{1}+k_{0}+6)
\sqrt{6k_{1}^{2}-12k_{1}-2k_{0}^{2}+4k_{0}}\phi}{6k_{1}-3k_{1}^{2}+k_{0}^{2}-2k_{0}}}-3(k_{0}+6-4k_{1})(k_{0}+\nonumber\\
&6-3k_{1})e^{\frac{(k_{0}+6)\sqrt{6k_{1}^2-12k_{1}-2k_{0}^{2}+4k_{0}}\phi}{k_{0}^{2}+6k_{1}-3k_{1}^{2}-2k_{0}}}+(k_{0}+6-5k_{1})(-3k_{1}^{2}+6k_{1}+k_{0}^{2}-2k_{0})c^{-2k_{1}+2}z_{h}^{-2k_{1}+6}\nonumber\\
&\Big(3k_{1}^{4}+(42-27k_{0})k_{1}^{3}+(-216+
46k_{0}+34k_{0}^{2})k_{1}^{2}+(-78k_{0}^{2}+44k_{0}+312-
11k_{0}^{3})k_{1}+k_{0}^{4}-\nonumber\\
&144+11k_{0}^{3}+40k_{0}^{2}-72k_{0}\Big)\Big].
\end{align}

\end{document}